\documentclass[aps,prb,twocolumn,showpacs,superscriptaddress,groupedaddress,floatfix]{revtex4-1}
\usepackage{latexsym}
\usepackage{graphicx}
\usepackage{dcolumn}
\usepackage{bm}
\usepackage{amssymb}
\usepackage{amsmath}
\usepackage[space]{grffile}

\usepackage{subfig}

\begin{document}

\newcommand{\niceref}[1] {Eq.~(\ref{#1})}
\newcommand{\fullref}[1] {Equation~(\ref{#1})}
\title{Entanglement properties of Floquet Chern insulators}

\date{\today}

\begin{abstract}
Results are presented for the entanglement entropy and spectrum of half-filled graphene following the switch on of a circularly polarized laser.
The laser parameters are chosen to correspond to several different Floquet Chern insulator phases.
The entanglement properties of the unitarily evolved wavefunctions are compared with the state where one of the Floquet bands is completely occupied.
The true states show a volume law for the entanglement, whereas the Floquet states show an area law.
Qualitative differences are found in the entanglement properties of the off-resonant and on-resonant laser.
Edge states are found in the entanglement spectrum corresponding to certain physical edge states expected in a Chern insulator. However, some edge
states that would be expected from the Floquet band structure are missing from the entanglement spectrum. An analytic theory is developed for the
long time structure of the entanglement spectrum. It is argued that only edge states corresponding to off-resonant processes appear in the
entanglement spectrum.
\end{abstract}

\author{Daniel J. Yates}

\author{Yonah Lemonik}
%\email{yl76@nyu.edu}

\author{Aditi Mitra}
%\email{aditi.mitra@nyu.edu}

\affiliation{Department of Physics, New York University, 4 Washington Place, New York, NY, 10003, USA}
\maketitle

\section{Introduction} 
Floquet topological insulators, a new kind of topological phase, have been attracting much attention in recent years \cite{Oka09,Inoue10,Kitagawa11,Lindner11,Kundu14,Torres14,Dehghani15a}. 
These phases are produced by periodically time-dependent Hamiltonians. 
By tuning the amplitude, phase, and frequency of the periodic drive myriad topological phases may be realized. 
Among Floquet topological insulators, effective time-reversal breaking Floquet Chern insulators (FCIs) 
have even been experimentally realized in cold-atoms~\cite{Esslinger14}  and optical wave-guides~\cite{Segev13}. 
These states contain unique $\pi$ edge states that have no equivalent in a static system~\cite{Rudner13}.

However, characterizing a driven system as topological is non-trivial, as the usual adiabatic arguments~\cite{Laughlin81} for topological properties cannot be applied. 
Naively, the topological nature of these systems could be deduced by analyzing the ground state of the Floquet Hamiltonian~\cite{Lindner11}, which we call the ``Floquet ground state'' (FGS). 
However the FGS has no general connection to the non-equilibrium state realized in experiment, and therefore it cannot be used to determine the existence of topological properties.

 In this work, we study the topological properties of FCIs using the entanglement spectrum (ES) \cite{Haldane08}. 
Entanglement statistics have already been 
demonstrated to detect ground states~\cite{TDMRGrev11,Eisert10,Eisert15}, critical states~\cite{Casini07,Metlitski09},  topological states~\cite{Levin06,Preskill06,Hamma13}, 
and universal exponents~\cite{Lemonik15}. 
The ES in particular is known to have edge states for topological states that mimic the edge states appearing in the 
physical boundaries of the same system~\cite{Fidkowski10}. 
As the ES is a function of the state of the system at a given time, it may be used to detect topological properties for a time-evolving state.  

We calculate the ES for four different phases of the FCI generated by a sudden turn on of the driving laser both numerically and analytically, 
and we complement these results with that for a slow turn on of the laser. We label states obtained this way, i.e., via unitary time evolution, as 
physical or true states,
and we compare the ES of these states with the ES of the FGS.
We present three main findings. First, the ES of the FGS correctly detects both the usual and $\pi$-edge states of the system. 
Second, in the case of a resonant laser, the FGS is qualitatively different from the physical state obtained from unitary time-evolution. 
Third, the ES in the physical resonant state does not show all the expected edge states. 
Therefore, the system driven by a resonant laser does not have the naively expected topological properties. These results
do not depend on how rapidly the laser was switched because, unlike for an off-resonant laser, for a resonant laser there is no adiabatic limit~\cite{Privitera16}. 
We also explicitly show this lack of adiabaticity of the resonant laser via the properties of the ES.

In Ref.~\onlinecite{Rigol14b} it was shown that a Chern number constructed out of unitarily evolved states is conserved. Thus 
if the initial state before the periodic drive was turned on had a zero Chern number, it would stay zero always. However, 
this Chern number is a property of the full density matrix and
is not related to physical observables directly. Rather, local physical observables
probe a local spatial region and therefore a reduced density matrix for which the arguments of Ref.~\onlinecite{Rigol14b} do not hold. 
Evidence that the Chern number of the full density matrix is not relevant also comes from the fact that 
that unitarily evolved states show a non-zero dc and ac Hall conductivity even for an initial state with zero Hall 
conductivity~\cite{Dehghani15a,Dehghani15b}, although this conductivity is smaller in magnitude than $C e^2/h$, 
as would be expected for the FGS. 
Thus our result that the ES of a unitarily evolved state reflects some of the topological properties of the FGS is consistent.

The paper is organized as follows. In section~\ref{sec1} the model is presented, and the topological properties of the FCIs of interest to us
are summarized. 
In section~\ref{sec2} the methods for obtaining the ES are described, while the results are presented and discussed in
section~\ref{sec3}. Many details are relegated to the appendices. 
Other than Figure~\ref{fig:slowFastTurnOn}, all other plots are for the sudden quench.
Appendix~\ref{app1}
contains results for the ES for the slow turn on of the laser. Appendix~\ref{app2}
gives the bulk occupation probabilities for the quench switch on protocol of the laser, as well as the 
projections of these occupation probabilities on the 
translationally invariant axis in order to highlight their relation to 
the bulk states of the ES.
Plots showing edge states, their chiralities, and decay lengths in the ES for FGS
and the unitarily evolved states are given in appendix~\ref{app3}.
A key equation whose solution yields the ES is derived in appendix~\ref{app4}, while 
analytic solutions for the edge states in the ES of the Dirac model are given in appendix~\ref{app5}, and they help to 
provide physical intuition for the more complex ES structure of the unitarily evolved state in the
presence of the laser.

%############### 3D OCCUPATION PLOTS ###########################

\section{System} \label{sec1}

We study the quench from the ground state of graphene at half-filling
to a time-periodic Hamiltonian corresponding to driving by a circularly polarized laser. The transport and optical
properties following such a quench has been extensively studied~\cite{Dehghani14,Rigol14b,Dehghani15a,Dehghani15b,Dehghani16}.
The Hamiltonian before the quench is that of a half-filled infinite sheet of graphene,
\begin{eqnarray}
&&H(t<0)=-t_h\sum_k\begin{pmatrix}c_{kA}^{\dagger}&c_{kB}^{\dagger}
\end{pmatrix}\nonumber\\
&&\times \begin{pmatrix}0&\sum_{i=1,2,3}e^{ia\vec{k}\cdot\vec{\delta}_i}\\\sum_{i=1,2,3}e^{-ia\vec{k}\cdot\vec{\delta}_i}
&0\end{pmatrix}\begin{pmatrix}c_{kA}\\c_{kB}\end{pmatrix},
\end{eqnarray}
where $a$ is the n.n. spacing, and $\vec{\delta}_1= \left(\frac{1}{2},\frac{\sqrt{3}}{2}\right);\vec{\delta}_2= \left(\frac{1}{2},-\frac{\sqrt{3}}{2}\right);
\vec{\delta}_3= \left(-1,0\right)$.
At $t=0$, the Hamiltonian is changed by substituting
$\left(k_x,k_y\right) \rightarrow \left(k_x+ A_0 \cos\Omega t, k_y- A_0\sin\Omega t\right)$,
so that the $H$ is now periodically time-dependent.
A state $|\Psi_k(t)\rangle$ of momentum $k$ evolves under this Hamiltonian for $t>0$ according to
\begin{eqnarray}
&&|\Psi_k(t)\rangle =
e^{-i \epsilon_{ka}t}|a_k(t)\rangle\langle a_{k}(0)|\psi_{k,\rm in}\rangle\nonumber\\
&&+ e^{-i \epsilon_{kb}t}|b_k(t)\rangle\langle b_{k}(0)|\psi_{k,\rm in}\rangle,
\end{eqnarray}
where $|\psi_{k,\rm in}\rangle$ is the initial state and $|a_k\rangle$, $|b_k\rangle$ are periodic in time and given by the Floquet-Bloch equation
\begin{equation}
\left[-i\partial_t -\epsilon_{ka} + H(t)\right]|a_k(t)\rangle = 0;
\label{Eq:FloquetBloch}
\end{equation}
and likewise for $|b_k(t)\rangle$. The $\epsilon_{ka,b}$ are the two quasi-energies, which are only defined up to integer multiples of $\Omega$. 
We take them to lie between $-\Omega/2 \leq \epsilon_{ka,b}\leq \Omega/2$ calling this range of $\epsilon_{ka,b}$ and $k$ the Floquet Brillouin zone (FBZ). 
We take $\epsilon_{ka} < \epsilon_{kb}$. The information of the initial state is encoded in the overlaps with $a$ and $b$ and may be quantified by the excitation density 
$\rho_{k,\rm down}-\rho_{k,\rm up}$ where 
\begin{eqnarray}
\rho_{k, \rm down}=|\langle a_{k}(0)|\psi_{k,\rm in}\rangle|^2,
\end{eqnarray}
is the occupation probability of the lower Floquet band and likewise for $\rho_{k,\rm up}$.

\emph{Topology} -- The wavefunctions at fixed $t$ and quasi-energies $\epsilon_{ka,b}$ may be interpreted as the band-structure of some underlying Hamiltonian,
\begin{equation}
H_{\text{eff}}(t) = \sum_{k,\sigma= a,b}\epsilon_{k\sigma}|\sigma_k(t)\rangle\langle \sigma_k(t)|,
\end{equation}
and we may define a ``ground state" of this Hamiltonian by completely filling the lower band
\begin{equation}
\left|{\rm FGS}\right> = \prod_k|a_k\rangle.
\end{equation}
It is found that for some choices of parameters, the corresponding Hamiltonian has a non-trivial Chern number and is 
therefore topological.
As in the static case, the Chern number is related to the existence of edge bands~\cite{Rudner13,Dehghani16}. In the FCI, there are two kinds of edge states:
``conventional'' edges that disperse through $\epsilon = 0$   and ``anomalous'' $\pi$ edges that disperse through $\epsilon = \pm\Omega/2$.
The Chern number is given by $C= N^{C} - N^A$ where $N^{C(A)}$ is the difference between the number of left-moving and right-moving conventional
(anomalous) edge modes. 

However, the natural physical question is whether the many-body state generically produced by experiment actually displays any topological properties. 
Note that despite appearances, there is no reason for the FGS to be produced in a generic experiment. As $\epsilon$ is periodic, there is no notion of a 
lowest-energy state and hence no notion of relaxation to a ground state. In fact, it is believed that generic disorder and interactions cause a 
driven closed system to reach infinite temperature~\cite{Lazarides15,Ponte15,Rigol14a}. Therefore we must consider the physical state as determined by time 
evolution under a reasonable experimental protocol, here a quench, both sudden and slow.

Having produced the physical state, we now must decide if it is topological. Here several standard arguments fail. 
As the system is time-dependent, there is no good notion of adiabatic flux threading~\cite{Laughlin81}. 
The application of a physical edge will qualitatively change the evolution of the system and therefore does not provide 
good information about the system in the absence of such an edge. As the application of generic disorder drives the system to infinite 
temperature, there is no notion of static localization. 

These problems are solved by using the entanglement spectrum (ES)~\cite{Haldane08}. The ES is calculated by imposing a fictitious boundary on 
the density matrix at a particular time.  The degrees of freedom outside the boundary are traced out, and the spectrum of the reduced density 
matrix is the ES. The fictitious entanglement boundary  functions similarly to a physical boundary, and it may host chiral edge states 
that can be used as evidence of topological properties. 

For a free system~\cite{Eisler2009}  the entanglement spectrum may be derived from the eigenvalues of the correlation matrix
\begin{equation}
C_{rr'} = \text{Tr}\left[\rho c^\dagger_r c_{r'}\right],
\end{equation}
where $c_r$ is the operator that annihilates an electron at site $r$, and the lattice sites $r$, $r'$ are restricted to lie in the sub-system of interest.
The eigenvalues $\epsilon$ of this matrix lie between $0$ and $1$ with a value of $0$ or $1$ representing an unentangled or pure state mode 
and $\epsilon = 1/2$ representing a maximally entangled mode. 

\begin{table}
\begin{ruledtabular}
\begin{tabular}{c|c|c|c|c|c|c}
Phase 	& $A_0a$ 	& $\Omega/t_h$ 	& Resonant? 	& $N^{C}$	& $N^{A}$& $C=N^{C}-N^{A}$ \\ \hline\hline
P$_1$		&	.5		&	10			&		x		&	1		&0	&1\\  \hline
P$_2$ 		& 	.5		&	5			& \checkmark 	& 1 		& -2&3	\\ \hline
P$_3$ 		& 	1.5 	&   5 			& \checkmark	& 1 		& 0	&1 \\ \hline
P$_4$ 		&	10		&  .5			&\checkmark 	&2			&2	&0
\end{tabular}
\end{ruledtabular}
\caption{\label{tab:phases}Summary of the four phases analyzed. Columns from left to right: Name of the phase, $A_0a$ the amplitude of the driving laser 
in units of the inverse lattice spacing, $\Omega/t_h$ laser frequency in units of the hopping strength, whether or not the laser is resonant with the 
static spectrum, $N^C$ number of usual edge modes of the FGS, $N^A$ number of anomalous $\pi$ edge modes, $C$ the  total Chern number.}
\end{table}

\section{Numerical Results} \label{sec2}
We consider four representative phases ($P_{1,2,3,4}$) of the FCI summarized in Table~\ref{tab:phases}. 
The first phase $P_1$  is the simplest case with an off-resonant laser $\Omega > 6 t_h$ and a single conventional edge mode. 
The other three  ($P_{2-4}$) are for resonant laser frequencies of $\Omega < 6 t_h$, where $\pi$ edge states may appear. 
 $P_2$ has one conventional edge mode and two counter-propagating $\pi$ modes so that the total Chern number is 3. 
 $P_3$ has a single conventional mode like $P_1$ but is produced by a high-amplitude resonant laser. Finally, $P_4$  is
an unusual topological phase where there are two conventional and two $\pi$ modes so that the total Chern number is zero. 
The occupation of these edge states following the laser quench in a system with boundaries was recently discussed in Ref.~\onlinecite{Dehghani16}. 
The occupation probabilities for the infinite system are discussed and plotted in appendix~\ref{app2}.

%############### MAIN PLOTS ###########################

\begin{figure}
\begin{tabular}{c}
{\includegraphics[width = 1.0\columnwidth]{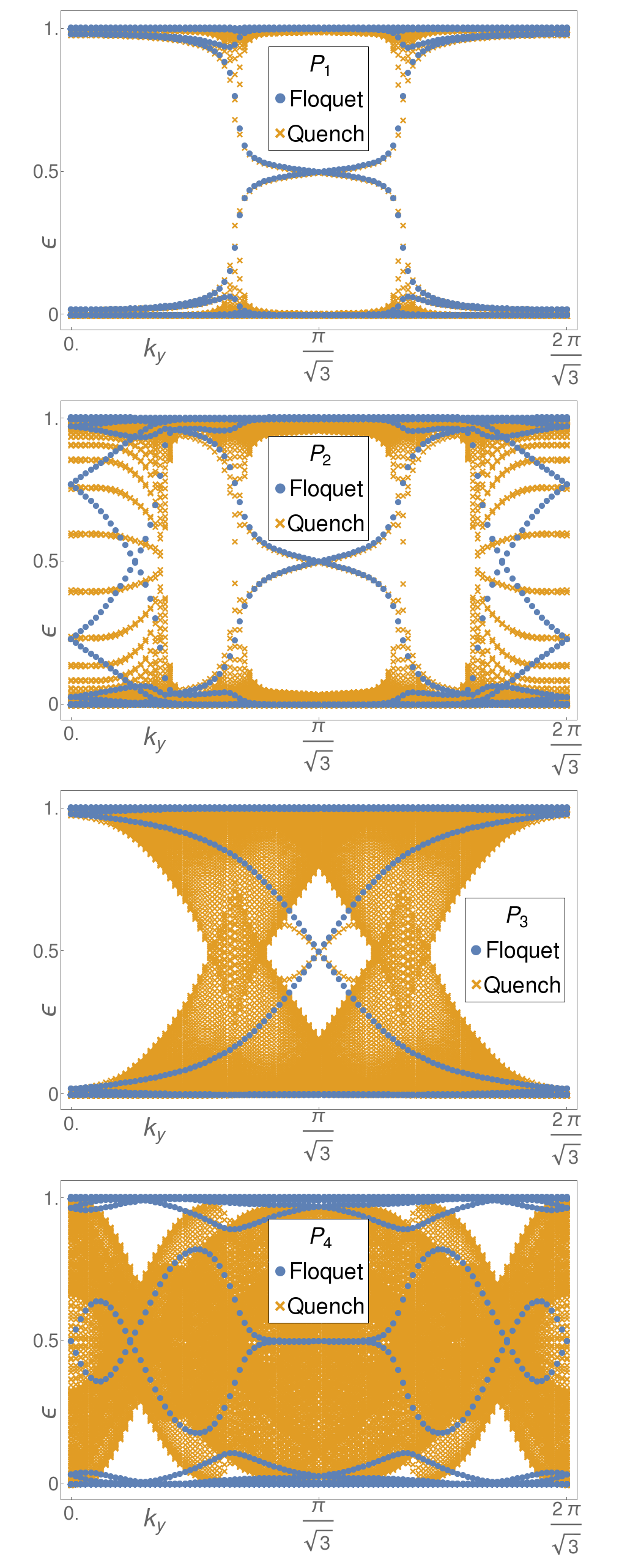}}
\end{tabular}
\caption{(Color online) Entanglement spectrum at $t=10,000$ laser periods after the quench, and compared with the ES for the Floquet ground state.
Edge modes correspond to  crossings at $\epsilon=1/2$. See appendix~\ref{app1} for slow quench.}
\label{fig4}
\end{figure}

We compute the long-time behavior of the ES and EE for the four phases $P_{1-4}$. The  time is fixed to be $10^4$ periods for the
sudden quench, whereas the results are at non-stroboscopic times for the slow quench (see appendix~\ref{app1}). 
The Floquet-Bloch wave-vectors are computed by expanding Eq.~\eqref{Eq:FloquetBloch} in a finite number of Fourier modes and solving the linear system. 
The result is used to construct $C_{rr'}$.

We take the entanglement boundary to be the zigzag edge of a strip of width $L$ sites so that the momentum $k_y$ along the strip remains 
a good quantum number. 
Thus, $C_{rr'}$ may be diagonalized at fixed $k_y$, and the results are plotted in Figure~\ref{fig4}. The results are compared with the ES of the FGS

\begin{figure}
\begin{tabular}{c}
%	\subfloat[t = 5000T]
{\includegraphics[width = 1.0\columnwidth]{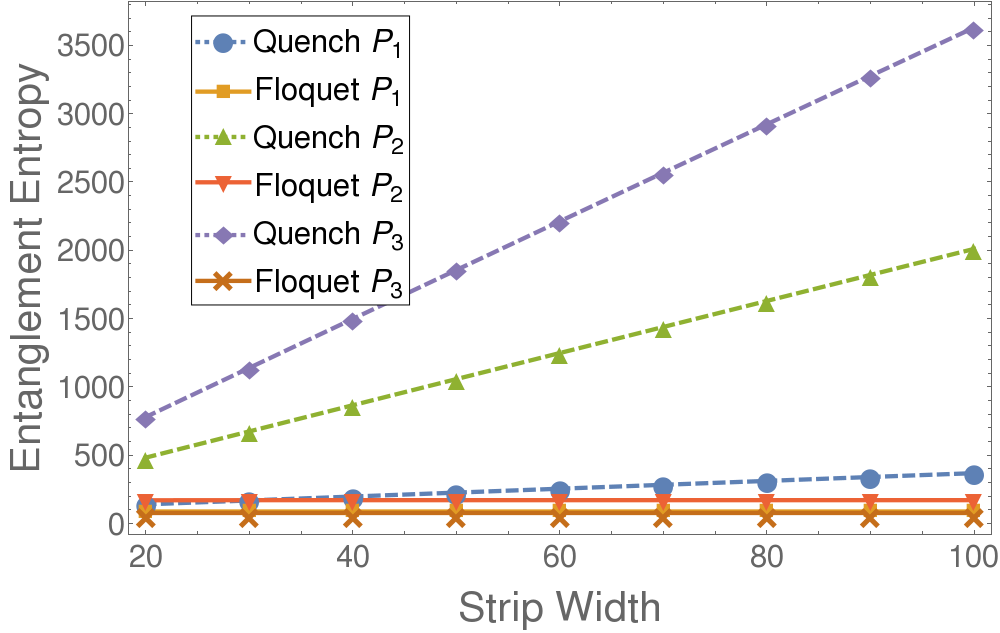}}
\end{tabular}
\caption{(Color online) EE of a semi-infinite strip at time $t =10,000$ laser periods after the quench, and compared with the EE for
a Floquet eigenstate. The state after the quench shows a volume law scaling by scaling linearly with strip-width, while the Floquet
eigenstate shows area-law scaling
by not depending on the width.}
\label{fig3}
\end{figure}

A single statistic that may be calculated from the ES is the entanglement entropy, given by
\begin{equation}
S_{A} = \sum\biggl[\epsilon\ln\epsilon + \left(1-\epsilon\right)\ln\left(1-\epsilon\right)\biggr],
\end{equation}
the sum being taken over all eigenvalues.
The results for the entropy are shown in Figure~\ref{fig3} for the FGS and the physical state.
 
The FGS state  shows area-law scaling, $S_A\propto L^0$. This is because the Floquet state is the ground state of $H_{\rm eff}$, which is local and gapped, 
and such states are known to show area law\cite{Wenbook,Eisert10}.
However the physical state shows volume-law scaling
($S_A\propto L$) as is expected of a generic state with ballistically propagating excitations. This behavior, therefore, does not provide detailed behavior about the phases.

The full ES for both states, namely the FGS and the physical state, for phases $P_{1-4}$ is shown in Figure~\ref{fig4} for the sudden
quench and Figure~\ref{fig:slowFastTurnOn} in Appendix~\ref{app1} for the slow quench. The figure is symmetric around $1/2$ as a consequence of
particle-hole symmetry. Let us first discuss the ES for the FGS. The ES shows a bulk of bands clustered near $0$ and $1$, with a gap between the two bands. 
Additionally, there are a small number of bands that disperse through $1/2$ that appear suggestively like chiral edge states.

On comparing with Table~\ref{tab:phases}, one finds that the number of edge states that cross the gap in the ES of the FGS is the same as the Chern number.
This means that when anomalous edge states appear, as in phases $P_{2,4}$, the chiralities of the edge states in the ES
reverses in such a way that the Chern number as calculated from the ES agrees with the Chern number of the phase. Thus for $P_2$, for example, there are 
3 chiral right-movers in the ES even though the anomalous modes are left moving at the physical edge. 
A similar reversal of chiralities is observed in $P_4$. 
Therefore, the ES ``understands'' that the $\pi$ modes contribute to the Chern number with opposite sign. 
While the ES is a snapshot at a particular time, at other times,
the location of the edge states changes, but their number and chirality are maintained so as to preserve the Chern number. 
Thus the phase $P_4$ can show odd behavior
due to the fact that it corresponds to $C=0$. This is highlighted in Figure~\ref{fig:P4Ent}, where within a laser period, 
the edge-modes can appear with canceling chiralities 
($t=0$ in figure), and totally disappear ($t=T/4$ in figure), both of these cases being consistent with $C=0$. 

Next we turn to the ES of the physical state. In the off-resonant case $P_1$, the Floquet and quench ES generally agree.  In the phases $P_{2-4}$,
the two appear drastically different. While the central crossing at $k_y=\pi/\sqrt{3}$ and the gap remain intact, a continuum of 
eigenvalues that pass through $\epsilon = 1/2$ appears in the quench case.
These continua of states cover the region where the two edge states of $P_2$ located at $k_y=\frac{\pi}{\sqrt{3}}(1 \pm 0.74)$ appear in the 
FGS spectrum and completely cover the gap in $P_4$. This picture qualitatively holds for a slow quench. As shown in 
Figure~\ref{fig:slowFastTurnOn}, the edge modes that are absent for
the sudden quench continue to be absent for a slower quench.
In the next section, we explain these results.
\section{Discussion} \label{sec3}

This behavior of the ES may be explained as follows. In the long-time limit, the contribution to the correlation function $C_{rr'}$ of the overlap between the 
upper and lower bands oscillates as $\exp i\phi(k)$ with $\phi(k) = |\epsilon_{ka} - \epsilon_{kb}| t-k (r-r')$. 
As $|r-r'|< L$, $\phi(k)$ changes rapidly with $k$ when $t \gg L$ and the overlap terms may be neglected.
Straightforward manipulation (see appendix~\ref{app4}) then gives $C_{rr'}\left(t\rightarrow\infty\right)= \frac{1}{2}\delta_{rr'} + D_{rr'}$ where,
\begin{align}
D_{rr'} &\equiv \int_{BZ}\!\!\!\! d^2k\,\,\delta\rho_k  M(k)
e^{i k\cdot(r-r')}\label{eq:Cinf},\\
M(k) &\equiv |a_k(t)\rangle\langle a_{k}(t)| - |b_k(t)\rangle\langle b_{k}(t)| \nonumber,\\
\delta\rho_k &=\frac{\rho_{k,\rm down}-\rho_{k,\rm up}}{2}.
\end{align}
If the size $L$  of the sub-region $A$ is taken to infinity, $D_{rr'}$ may be diagonalized by Fourier transformation, and its eigenvalues may be read off
from Eq.~\eqref{eq:Cinf} as $\pm \delta\rho_k$ with eigenvectors $|a_k(t)\rangle$ and $|b_k(t)\rangle$. Therefore we may think of $D$ as a Hamiltonian with
two bands, where the gap between the two bands is $2\delta\rho_k$, and the wavefunctions have some topological properties.
Note that the smaller the quench-induced excitation density is, the larger is the gap $2\delta\rho_k$ in the ES.

The problem of finding the spectrum of $D$ at finite but large $L$ is then akin to the problem of finding the spectrum of a Hamiltonian in a finite geometry.
We expect two kinds of states: bulk states and edge states. The bulk states should have the ``energies'', $\delta\rho_k$, but where $k$ is quantized
in the finite direction. Therefore, the ``bulk'' states can be produced by projecting the occupation number  onto the $k_x = 0$ plane (the translationally
invariant direction); see Figure~\ref{fig:Proj} in appendix~\ref{app2}.

In addition to these bulk states, there may be edge states.
If the Floquet bands are topological, that is, if the Chern number is non-zero, \emph{and} $\delta\rho_k$ is non-zero,
then the topological protection of edge modes of $D$ applies and there will necessarily be edge modes in the ES. This property is topological 
in the sense that a small deformation of the state, which is equivalent to a deformation of $\delta\rho_k$ and $M(k)$, cannot change the net number of edge modes.
For the FGS, where $2\delta\rho_k =1$ by definition, there will exist edge modes that agree with the Chern number of the bands.
This also appears to be the case in $P_1$, where $\delta\rho_k=0$ only at isolated points, and the edge modes of the FGS and physical states agree,
although perhaps the issue is more delicate when the entanglement cut is very rough so that momentum $k_y$ is not conserved.

That these states are indeed topological can be seen in the numerics. Their number does not scale with $L$ and their crossings are not lifted 
by varying the phase of the driving laser. Moreover, a direct analysis of the wavefunctions associated with the eigenvalues shows that they are chiral and 
located on the edge. Figures~\ref{fig:EdgeF12} and ~\ref{fig:EdgeF34} 
plot the wavefunctions for the edge states for the FGS, while Figures~\ref{fig:EdgeQ12} and~\ref{fig:EdgeQ34} plot the
same for the true state, and Figure~\ref{fig:edgedecay} highlights the decay length of the edge states.

If $\delta\rho_k = 0$, then this is equivalent to a gap closing in the Hamiltonian, and topological protection arguments no longer hold. 
Indeed, we find that the edge states do not coexist with the bulk states in the numerics.
Since a resonant process will co-occur with a population inversion (see Figure~\ref{fig13}), we do not expect edge states to be
protected in the resonant case.  Further, since $\delta\rho_k=0$ precisely where the resonant laser causes band crossings, the
resulting edge modes are not robust at all. 
Indeed, this is what is seen in the phase $P_2$ where the two edge modes seen in the Floquet bands at the boundary of the FBZ
do not appear in the ES of the true state. This is not an artifact of the quench protocol, and similar population inversions appear in the slow turn on of the 
resonant laser (see appendix~\ref{app1}).

Although the edge modes lose topological protection when $\delta\rho_k = 0$,
there may still be non-protected edge states. Since we study a perfectly clean system with translational invariance in the $y$ direction,
an edge mode will appear under the much weaker condition that $\delta\rho_k\neq 0$ for the
same $k_y$ at which the edge mode appears. Therefore, we find that the central mode appears in $P_2$ and $P_3$ even though
$\delta\rho_k$ is zero over a large arc in the BZ.
We expect that this edge mode would disappear for other geometries.  For $P_4$, the high-amplitude laser produces population inversions throughout the
BZ, and there are no edge states whatsoever in the physical state.
For phase $P_1$, on the other hand, one has a robust gap, with gap closings only at some special points (rather than large arcs). These excitations
can be made even smaller for a slow laser switch on, as for the off-resonant case, an adiabatic theorem can hold.  

This provides a complete description of the entanglement spectrum at long times after a Floquet quench.
The key quantity that determines the applicability of the Chern number calculation to the edge modes of
the quench is the population difference $\delta\rho_k$. 

\section{Conclusions}
It is interesting to ask how the results of the ES compare with physical observables.  In Ref.~\onlinecite{Dehghani16}, the occupation
of the edge states in a system with boundaries was studied, and it was found that the edge state for the off-resonant case $P_1$
was occupied at a low effective temperature. In contrast, only one set of edge modes in the phase $P_2$ was occupied at a low effective 
temperature, while the two new sets of  $\pi$-edge modes appearing due to resonant band crossings were occupied at a high effective temperature
and coexisted with bulk excitations.
The dc  conductivity for these phases was studied in Ref.~\onlinecite{Dehghani15a}, where it was found that 
the conductance was very close to the maximum, $\sim.8e^2/h$, for $P_1$, and it persisted at this value for $P_2$ even though new
edge modes appear for $P_2$. This observation implies that the high effective temperature of the resonant edge modes prevent them from 
contributing much to transport, in contrast to the off-resonant edge mode. 

Thus our results show that the study of the ES is a useful way to understand the topological properties of a periodically driven system.
An important direction of research would be to extend this analysis to interacting Floquet topological insulators.

{\sl Acknowledgements:}
This work was supported by the US Department of Energy,
Office of Science, Basic Energy Sciences, under Award No.~DE-SC0010821.

\begin{appendix}

%################# NUMERICALLY PROPAGATED PLOTS ############################
\begin{figure}
\begin{center}
\includegraphics[width = 1.0\columnwidth]{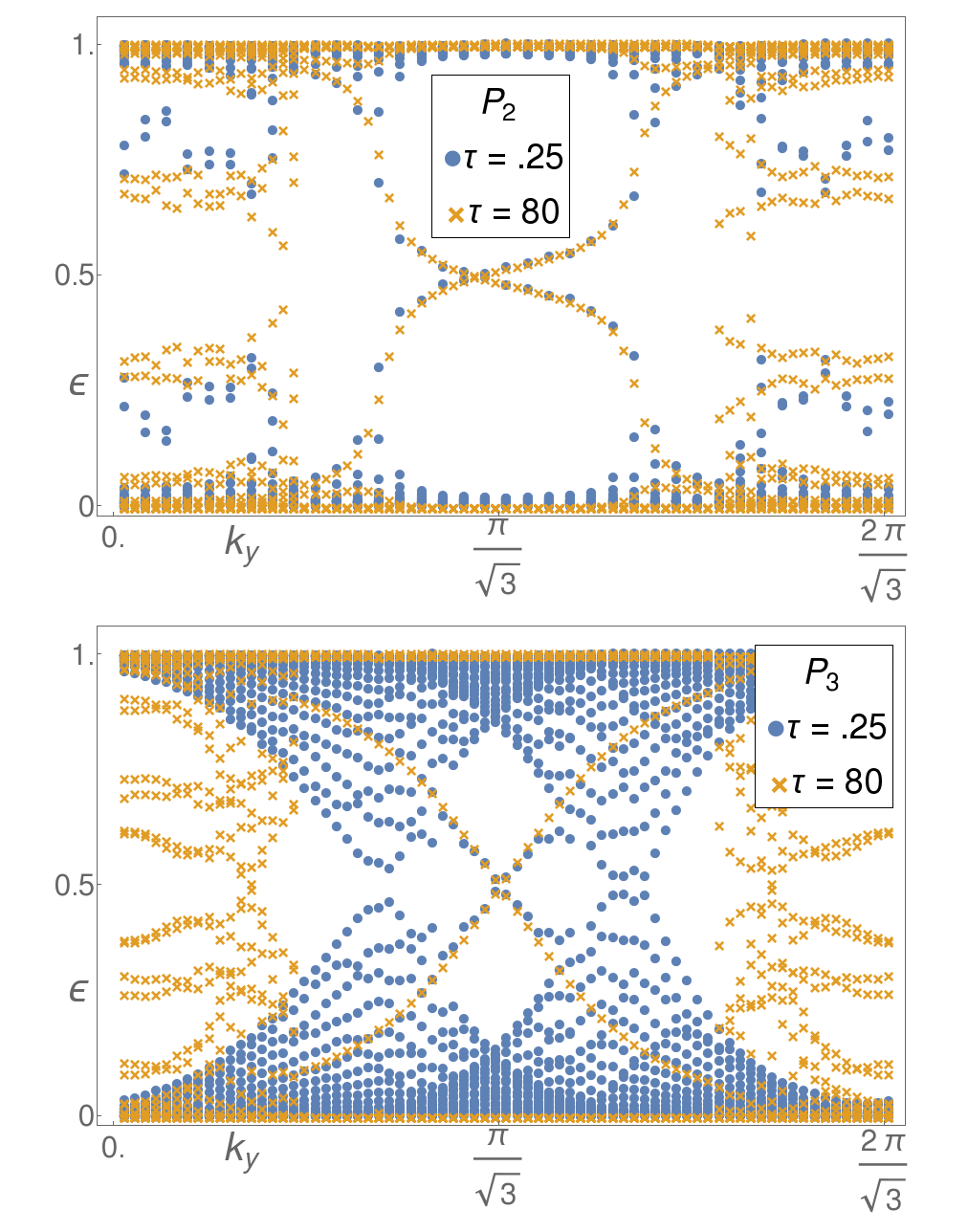}
\end{center}
\caption{(Color online) Entanglement spectrum (ES) resulting from a numerically propagated state for a strip width of 20 cells.
The time constants $\tau$ correspond to the ramp speed of the laser light, of the form
$\frac{A_0}{2}\biggl[\tanh(\frac{t}{\tau}) + 1\biggr]$.
ES shown at time $t=700.5 T$ for
fast ramp, and at time $t=800.5T$ for slow ramp, where $T$ is the laser period.
}
\label{fig:slowFastTurnOn}
\end{figure}

\section{Results for the ES for the slow turn on of the laser}\label{app1}
To generate occupation probabilities for the slowly turned on laser, the Schr\"odinger equation at fixed $k_y$
is solved by implementing the Commutator Free Exponential Time method following Ref.~\onlinecite{Alvermann12}. The occupation probabilities
and entanglement spectrum are calculated from these wavefunctions. Due to the computational cost of this procedure, the system width was taken to be
$20$ sites.

The results for the ES shown in Figure~\ref{fig:slowFastTurnOn}, are in qualitative agreement with the quench protocol discussed in
the body of the paper. Band inversions appear near momenta where the laser is resonant with the band gap. These band inversions are
accompanied by a high bulk excitation density, leading to a closing of the entanglement gap, and ruining the topological protection
of edge states.

Figure~\ref{fig:slowFastTurnOn} shows data for two different laser switch-on rates for the two resonant phases $P_{2,3}$. $P_2$
has three pairs of chiral edge states in the ground state (FGS) of the Floquet Hamiltonian. However, we find that for all switch-on rates
of the laser, two of the three pairs of edge states (located at $k_y=(1\pm 0.74)\pi/\sqrt{3}$) are absent from the ES of the unitarily evolved states.
Only the central edge state located at $k_y=\pi/\sqrt{3}$ survives.

For phase $P_3$, there is only one pair of chiral edge states in the Floquet ground state, and this is visible in the ES of all
the unitarily evolved states as well. However this edge state coexists with bulk excitations that survive both for a fast and slow laser quench.

Paradoxically, a slower laser switch on creates more bulk excitations when the laser frequency is
resonant. We understand this as follows. A large-amplitude laser can
modify the band structure considerably, making the bands flatter. Thus a resonant laser can become effectively off-resonant
 when the amplitude of the laser becomes too large. Thus when a laser amplitude is switched on slowly, (while its
frequency is kept unchanged), since for a longer period the effective amplitude of the laser is smaller for a slower switch
on than a faster switch on, more resonant excitations are created for the slower switch on
as the resonance condition is obeyed for a longer duration of time. That is the reason why in Figure~\ref{fig:slowFastTurnOn},
there are more bulk excitations for the slower switch on than for the faster switch on.  

The only reason why the central edge state at $k_y=\pi/\sqrt{3}$ is unaffected by these bulk excitations
is because our entanglement cut is smooth and momentum $k_y$ is conserved. This prevents hybridization between the central
edge state and the bulk excitations, since the latter are located at other values of $k_y$. However a rough entanglement cut will
wipe out even this central edge state.

\section{Occupation probabilities and their projection on $k_x=0$ axis}\label{app2}
\begin{figure}
\begin{tabular}{c}
	\subfloat[$P_1=\left(A_0a,\Omega/t_h\right)=\left(.5,10\right),C=1$. Inset: Solid line indicates conventional BZ,
	dashed line shows the BZ we consider.]{\includegraphics[width = 1.0\columnwidth]{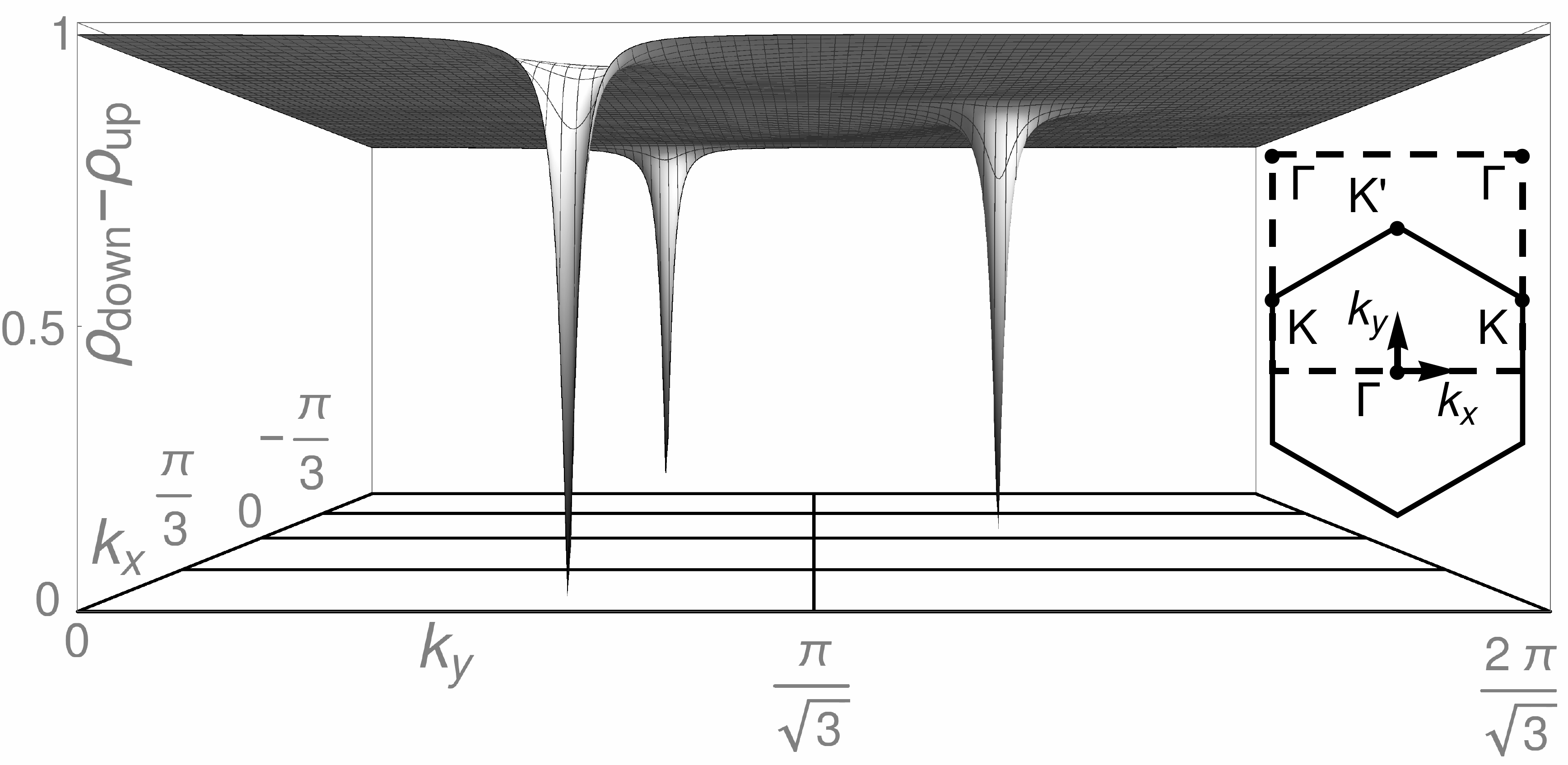}}\\
	\subfloat[$P_2=\left(A_0a,\Omega/t_h\right)=\left(.5,5\right),C=3$]{\includegraphics[width = 1.0\columnwidth]{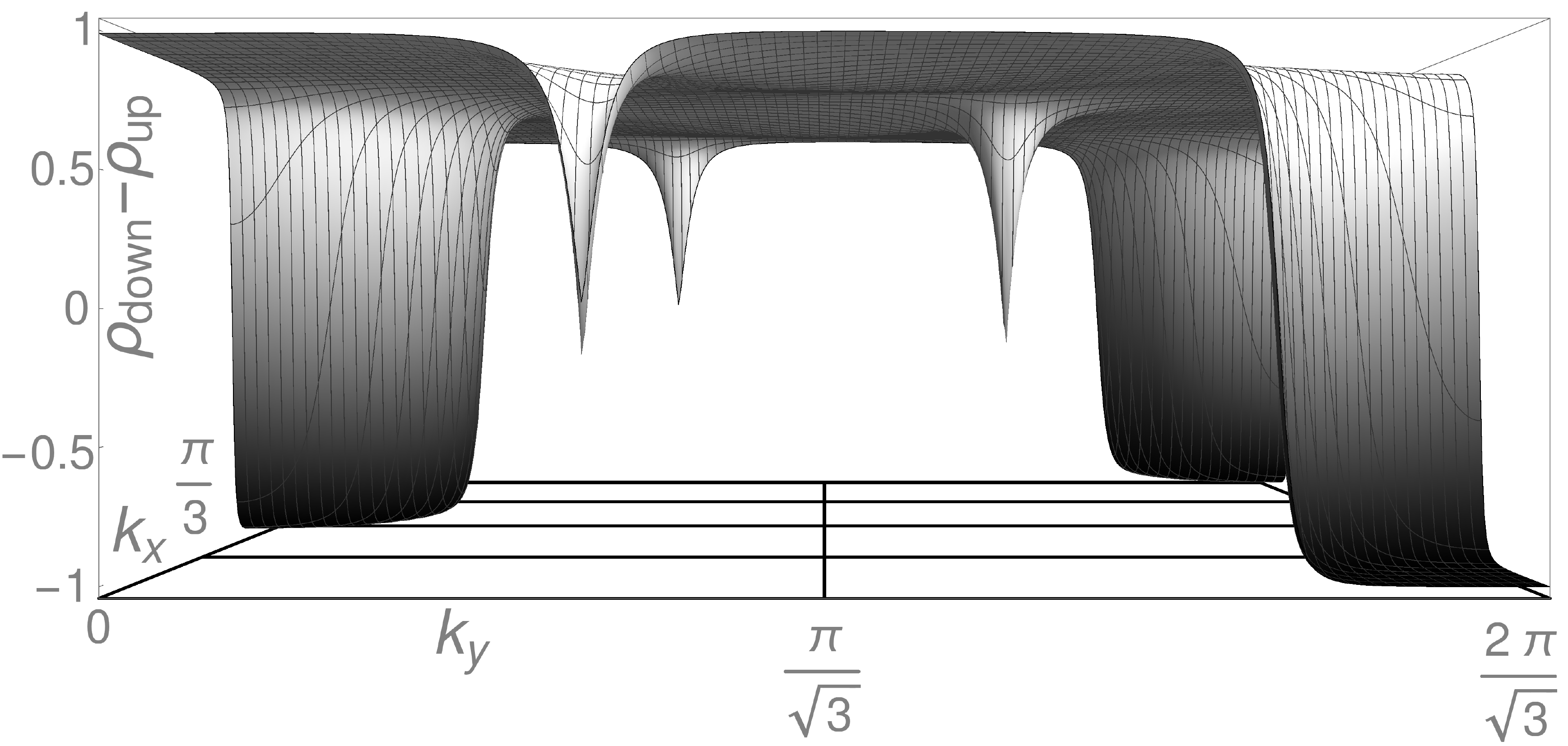}}  \\	
\end{tabular}
\caption{(Color online) Occupation probability for a laser quench for the phases $P_{1,2}$ at half-filling. $\rho_{k,\rm down}+\rho_{k,\rm up}$=1.}
\label{fig:drhoP12}
\end{figure}

The occupation probabilities of the Floquet bands for phases $P_{1,2}$ are shown in Figure~\ref{fig:drhoP12}, and those for the phases $P_{3,4}$ are shown in
Figure~\ref{fig:drhoP34}. Except for $P_1$, all the phases correspond to a resonant laser.

For $P_1$, the quasi-energy of the Floquet Hamiltonian is similar in appearance to static graphene with the
important modification of becoming gapped at K and K'. Thus, the occupation probabilities follow intuitively from the static picture, and the 
sharp spikes in Figure~\ref{fig:drhoP12} correspond to excitations around the K and K' points. 
The laser being off-resonant is equivalent to the statement that the periodicity of the quasi-energy 
does not play a role since corresponding boundaries in the FBZ are far from the maximum and minimum of the bands. 

Both $P_{2,3}$ have lasers of the same frequency,
but the amplitude of the laser for $P_3$ is larger than that for $P_2$. In $P_2$ there is a population inversion at $\Gamma$ in addition to the excitations at 
K and K'. This is simply the resonance condition, as now the frequency of the laser is such that it connects the peak and trough of the static graphene bands. 
In the quasi-energy picture, this translates to the original static bands extending beyond the FBZ and thus forming quasi-energy bands that avoid one another 
and  bend away at the boundary of the FBZ. 

A large amplitude laser modifies the effective band structure to such
a degree that a resonant laser becomes effectively off-resonant. The bands are drawn toward the center of the FBZ, away from the boundaries.
It is for this reason that $P_3$ has fewer bulk excitations than $P_2$, despite the two
phases sharing the same laser frequency. 
The flattening of the bands explains the broadening of the excitations at the K and K' points.

For phase $P_3$, even though the laser is resonant, the excitation density around $k_y=\pi/\sqrt{3}$, where the central edge mode in the ES appears,
is still fairly low.  
The laser creates pockets of excitations in regions symmetrically located around the central edge mode. Momentum conservation
prevents the central edge mode from mixing with these pockets of bulk excitations. 

The amplitude of the laser for
$P_{4}$ is much larger than all the other phases, and its frequency is much smaller as well.
As a consequence, the $P_4$ phase is highly excited, with the two Floquet bands
being almost equally occupied.  

Figure~\ref{fig:Proj} is the projection of the occupation probabilities onto the $k_x=0$ axis.
These projections are generated by selecting constant $k_x$ slices of the occupation probabilities that correspond 
to the modes that satisfy the boundary conditions of the strip. These slices are then superimposed on top of one 
another to give the projection image.

As explained in the main text, these projected plots reproduce the bulk states of the ES. 
There are some small discrepancies between the projection plot and the 
true spectrum. As seen in Figure~\ref{fig4}, the bulk excitations on either side of the central edge state 
in the quenched $P_2$ phase are smooth, but the analogous bands constructed from 
Figure~\ref{fig:Proj} appear ragged and cross $\epsilon \sim 1/2$, where they do not in the former. 
We explain this difference below.

As the width is increased, we first expect the bulk excitation bands of the ES to take on the rough appearance as predicted in the
projection plot. The larger width will also lead to more bands coalescing towards $\epsilon = 1/2$. Thus in the very large limit, we expect a 
continua of excitations on both sides of the central edge state of the quenched $P_2$ phase. 
The reason for the differing appearances in the large, but not extreme width limit is due to the bulk bands having residual knowledge of the 
edges and thus undergoing a smoothening and repulsion procedure, as one expects for perturbations. Thus as the width is increased, this smoothening will 
diminish and bands from both Figures~\ref{fig4} and~\ref{fig:Proj} will agree. The reason the repulsion argument does not extend to the edge states is precisely 
due to their local nature and exponentially small overlap. 

In the $P_3$ phase, $k_x$ projections originate from  occupation probabilities that have a relatively smooth structure throughout the FBZ, 
as seen in Figure~\ref{fig:drhoP34}. Thus we do not run into this problem for the $P_3$ phase because the projection already creates a continuum that agrees with 
the spectrum in Figure~\ref{fig4}. However, for the $P_2$ phase, the probability occupations 
have sharp, nearly vertical transitions that lead to discrete bands that single out and amplify this otherwise small feature. The continuum of excitations will only be
created by the projection plot for $P_2$ if the strip width is so large that the spacing between $k_x$ states is narrow enough to allow for many states to be
selected along the cliff faces (see Figure \ref{fig:drhoP12}) found in the occupation probability of the $P_2$ phase.

\section{Edge states in the ES, their chirality, and decay lengths} \label{app3}

The entanglement spectrum for phase $P_4$ is shown in Figure~\ref{fig:P4Ent}. The Floquet state shows the behavior expected of a $C=0$ Floquet Chern insulator.
The ``bulk'' bands are clustered around $\epsilon = 0$ and $\epsilon =1$. There are edge states seen around $\epsilon = 1/2$,
however these do not have a net chirality. This can be seen either from direct inspection of the wavefunctions in Figure~\ref{fig:EdgeF34},
or by noting that the
edge states do not connect with the bulk bands. Therefore, these edge states are not topologically protected, and they should disappear under disorder.
In fact Figure~\ref{fig:P4Ent} shows that the edge states appear at certain times in the laser cycle, and they disappear for certain other times.
This does not happen for phases $P_1$, $P_2$, and $P_3$; when probed at times away from an integer number of laser periods, the precise 
location of the edge states is shifted in the ES, but the results for the number and chirality of the edge states that are visible remain the same. 

\begin{figure}
\begin{tabular}{c}
	\subfloat[$P_3=\left(A_0a,\Omega/t_h\right)=\left(1.5,5\right),C=1$]{\includegraphics[width = 1.0\columnwidth]{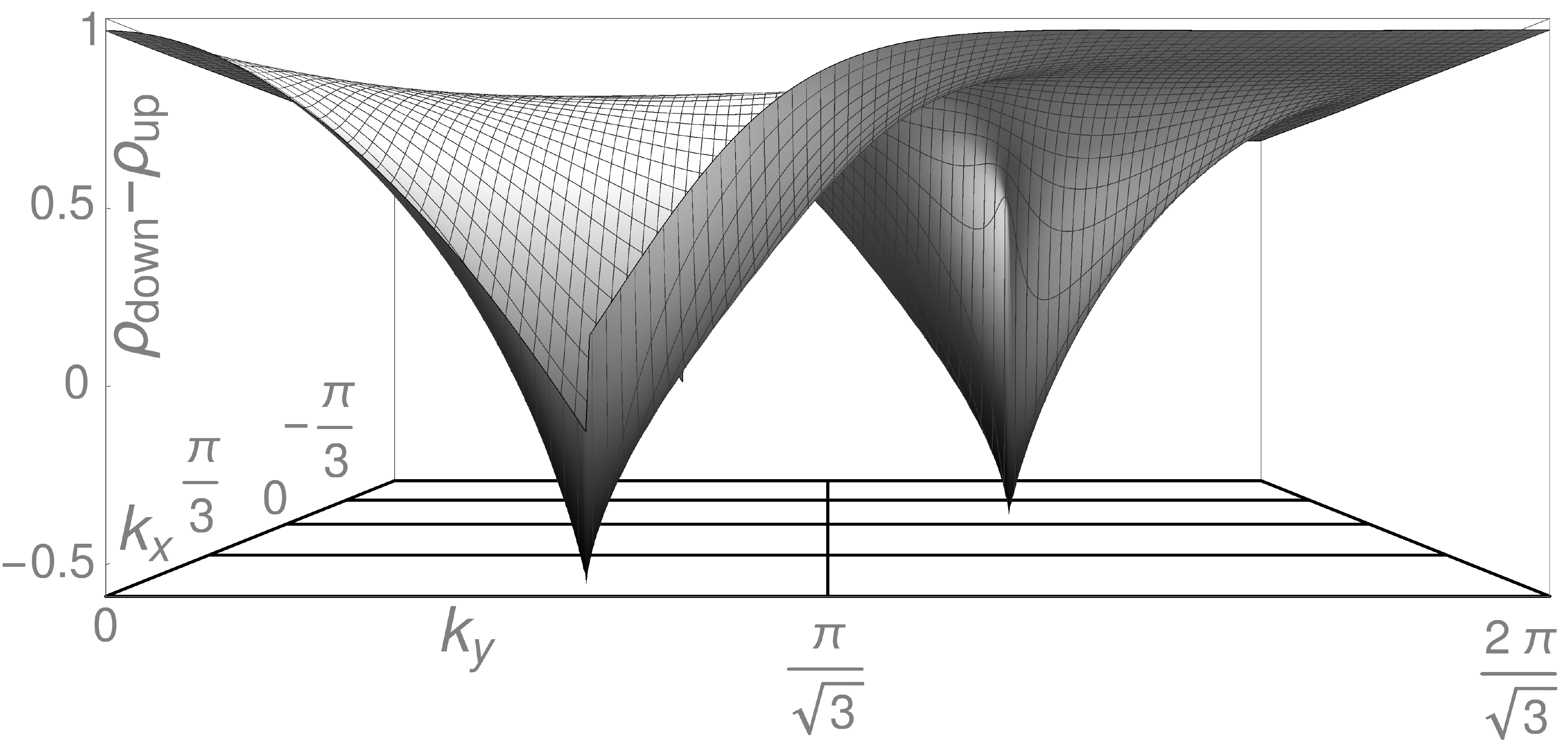}}\\
	\subfloat[$P_4=\left(A_0a,\Omega/t_h\right)=\left(10,.5\right),C=0$]{\includegraphics[width = 1.0\columnwidth]{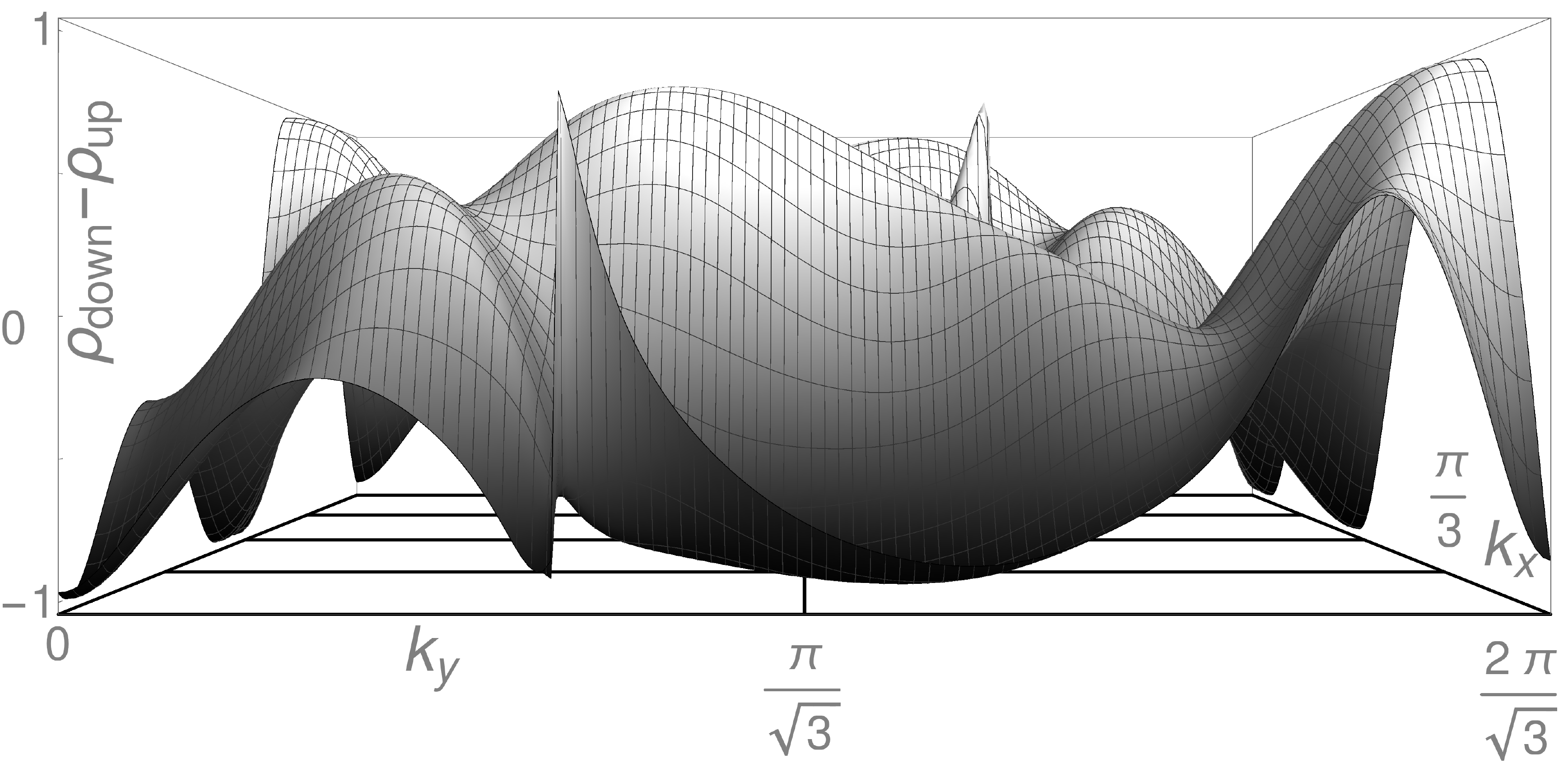}}
\end{tabular}
\caption{(Color online) Occupation probability for a laser quench, for the phases $P_{3,4}$ at half-filling. 
$\rho_{k,\rm down}+\rho_{k,\rm up}$=1. 
The $P_4$ phase is a high amplitude, resonant laser, so the occupation of each Floquet band is roughly $1/2$ throughout the FBZ.}
\label{fig:drhoP34}
\end{figure}

\begin{figure}
\begin{tabular}{c}
	\subfloat[$P_1=\left(A_0a,\Omega/t_h\right)=\left(.5,10\right),C=1$]{\includegraphics[width = 1.0\columnwidth]{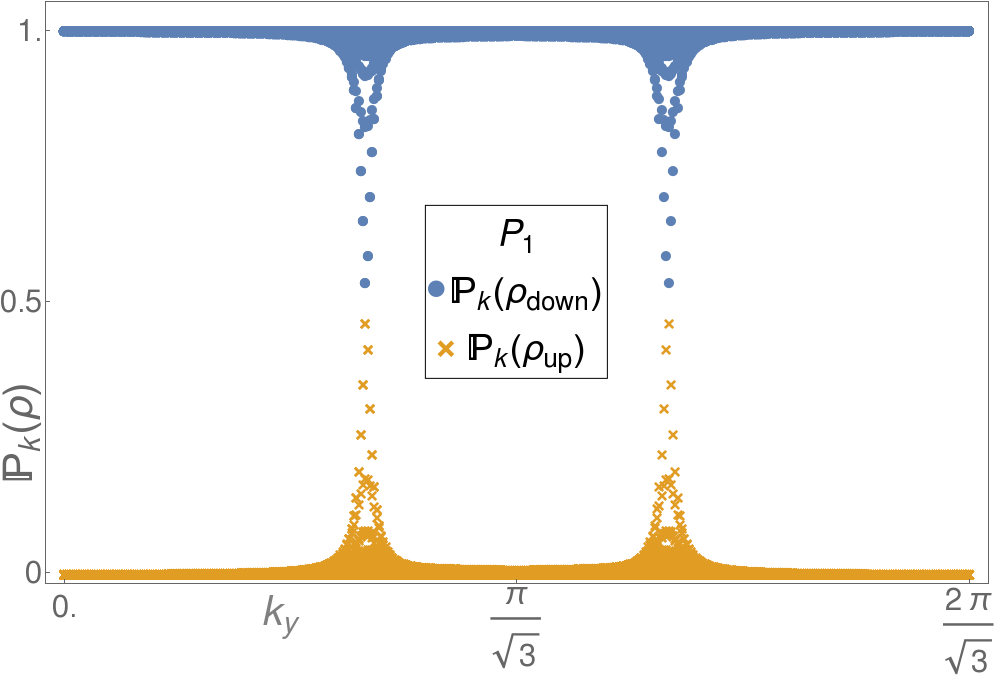}} \\
	\subfloat[$P_2=\left(A_0a,\Omega/t_h\right)=\left(.5,5\right),C=3$]{\includegraphics[width = 1.0\columnwidth]{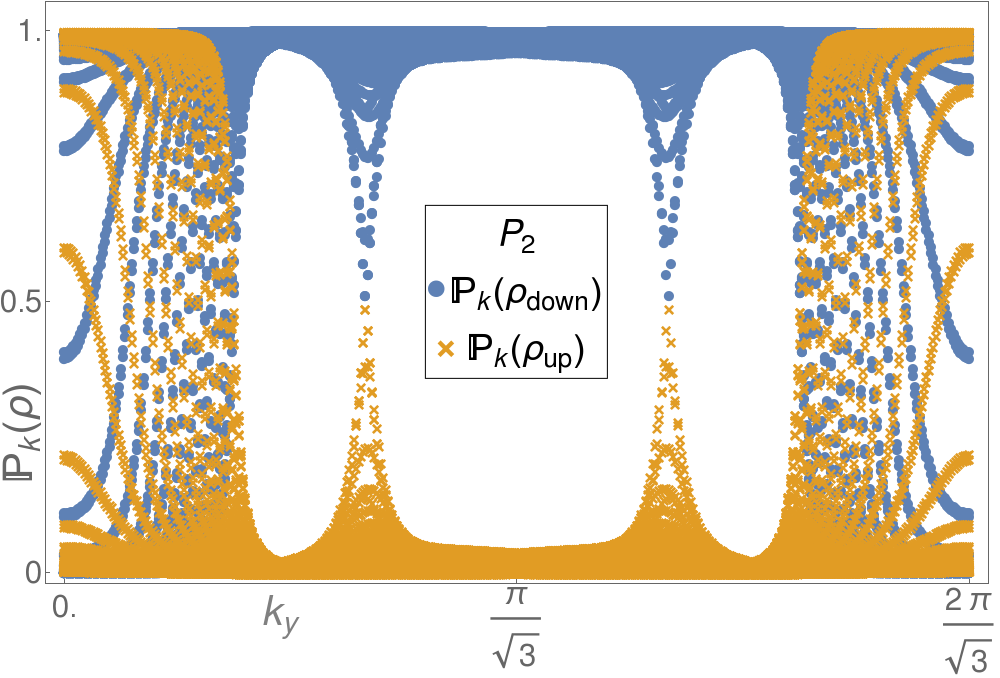}}  \\
	\subfloat[$P_3=\left(A_0a,\Omega/t_h\right)=\left(1.5,5\right),C=1$]{\includegraphics[width = 1.0\columnwidth]{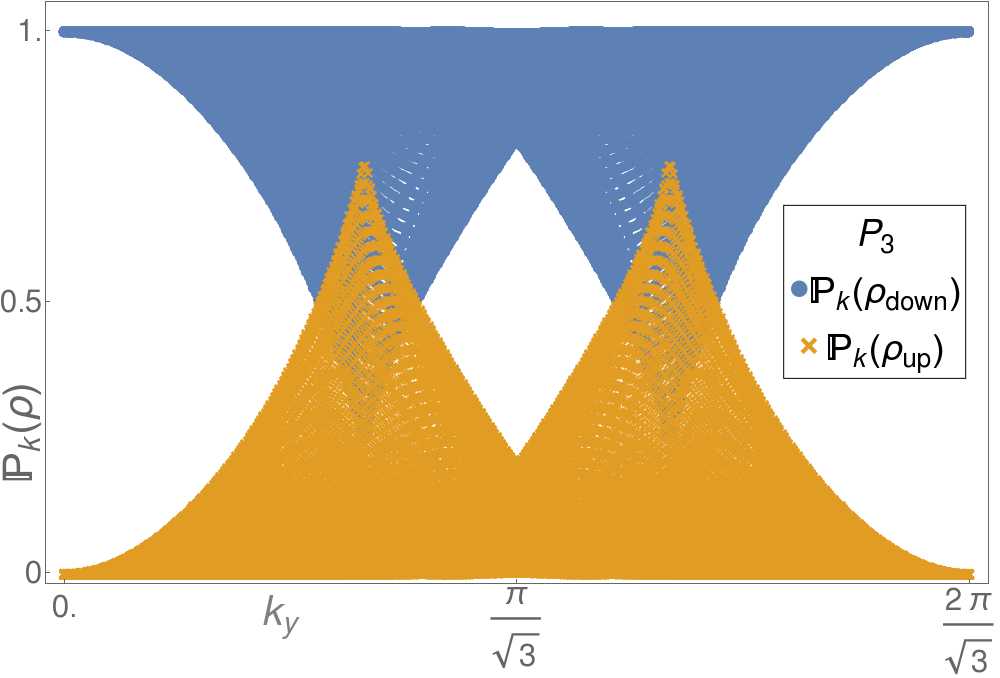}}
\end{tabular}
\caption{(Color online) Projection of the occupation probabilities of the lower and upper Floquet band onto the $k_y$ axis agrees well
with the ES for the bulk states in Fig.~\ref{fig4}. Edge states have to be accounted for separately.}
\label{fig:Proj}
\end{figure}

\begin{figure}
\begin{tabular}{c}
	\subfloat[$P_4=\left(A_0a,\Omega/t_h\right)=\left(10,.5\right), C = 0$]{\includegraphics[width = 1.0\columnwidth]{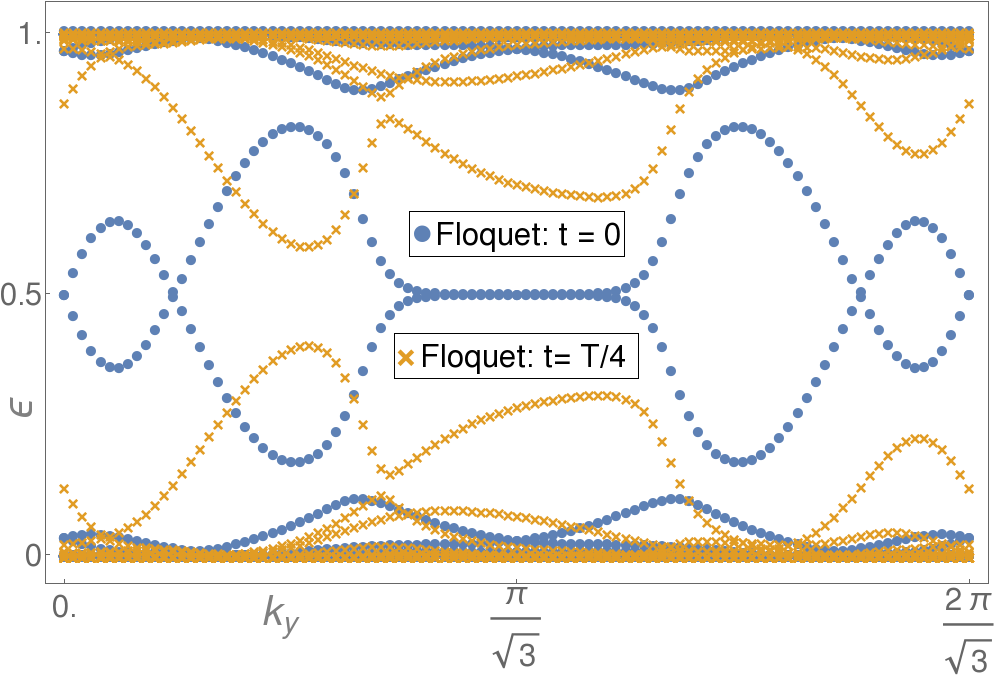}}
\end{tabular}
\caption{(Color online) ES for FGS for topological phase $P_4$. The edge modes at
certain times within one period of the laser can disappear. This is reflected by the absence of
crossings at $\epsilon=1/2$ for $t=T/4$.}
\label{fig:P4Ent}
\end{figure}

The true quench state for phase $P_4$ does not show any edge states, and the bulk bands approximately fill the space (see figure in main text).
This is in agreement with the occupation number difference shown in Figure~\ref{fig:drhoP34}(b),
which is found to vary widely between $-1$ and $1$ and crosses
$0$ on large arcs through the BZ. From the perspective of the entanglement and occupation number, therefore,
the true quench state for $P_4$ appears similar to a thermal state.

\begin{figure}
\begin{tabular}{c}
	\subfloat[$P_1=\left(A_0a,\Omega/t_h\right)=\left(.5,10\right),C=1$]{\includegraphics[width = .9\columnwidth]{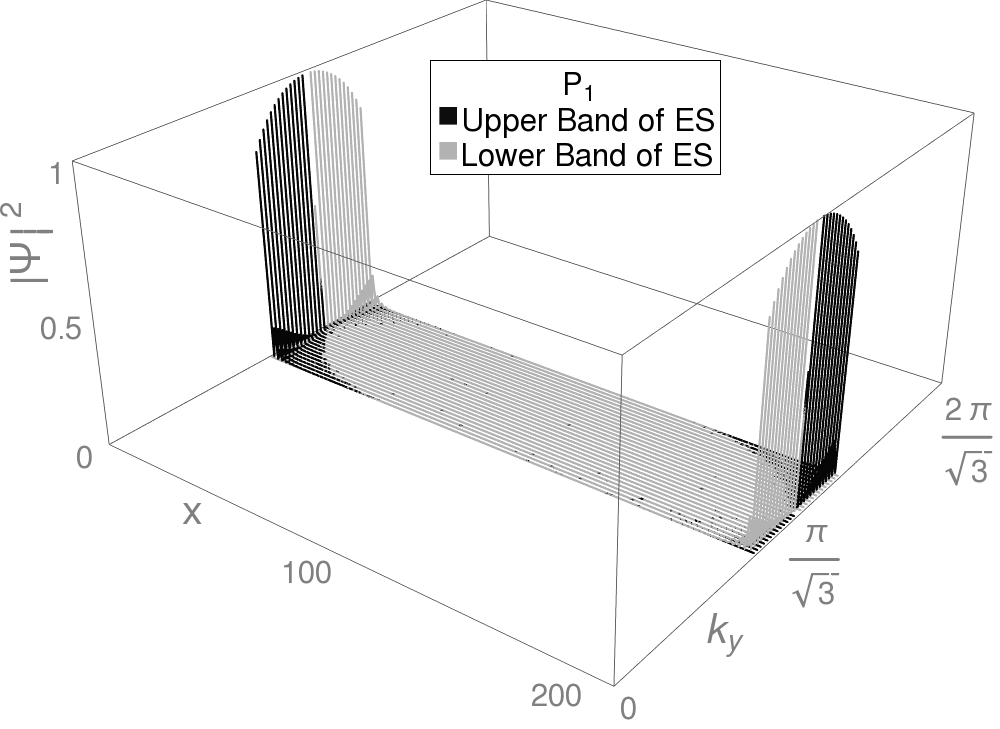}}\\
	\subfloat[$P_2=\left(A_0a,\Omega/t_h\right)=\left(.5,5\right),C=3$]{\includegraphics[width = .9\columnwidth]{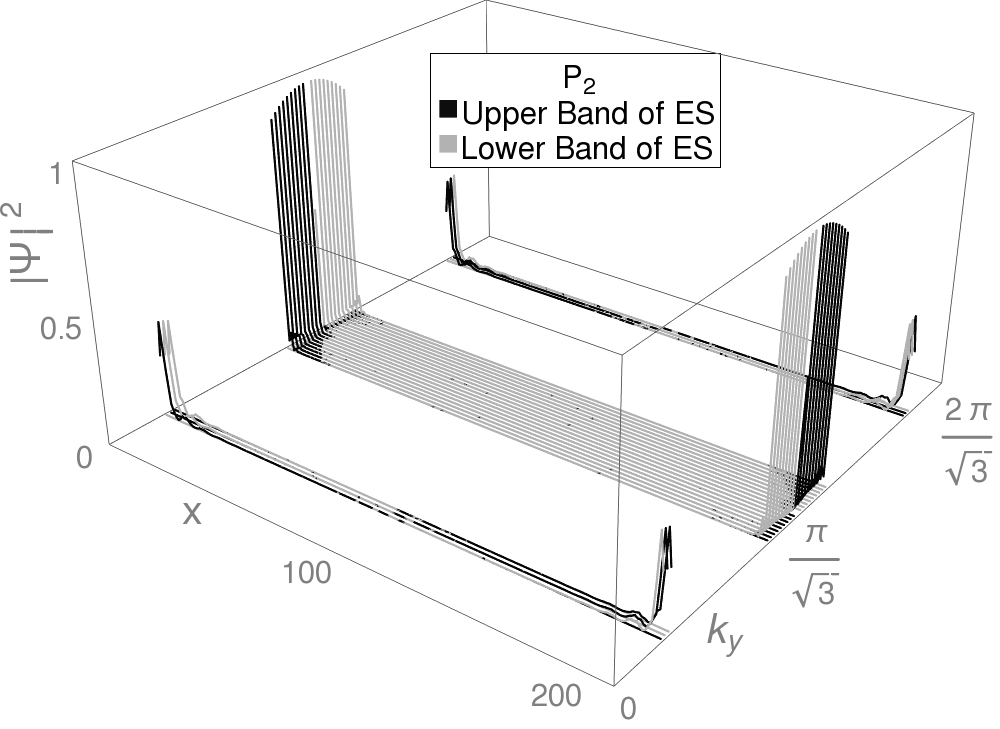}}
\end{tabular}
\caption{(Color online) Edge modes of the Floquet state for the phases $P_{1,2}$.  Upper (lower) band corresponds to entanglement eigenvalues
$\epsilon \geq 1/2$ ($\epsilon \leq 1/2$). Only states corresponding to $\epsilon$ within a small window of $1/2$ are shown, to highlight those 
that disperse through $1/2$. The $P_1$ phase contains a left mover on $x=0$ edge and a right mover on the $x = 200$ edge.
The $P_2$ phase contains three left movers on the $x=0$ edge and three right movers on the $x = 200$ edge.}
\label{fig:EdgeF12}
\end{figure}

\begin{figure}
\begin{tabular}{c}
	\subfloat[$P_3=\left(A_0a,\Omega/t_h\right)=\left(1.5,5\right),C=1$]{\includegraphics[width = .9\columnwidth]{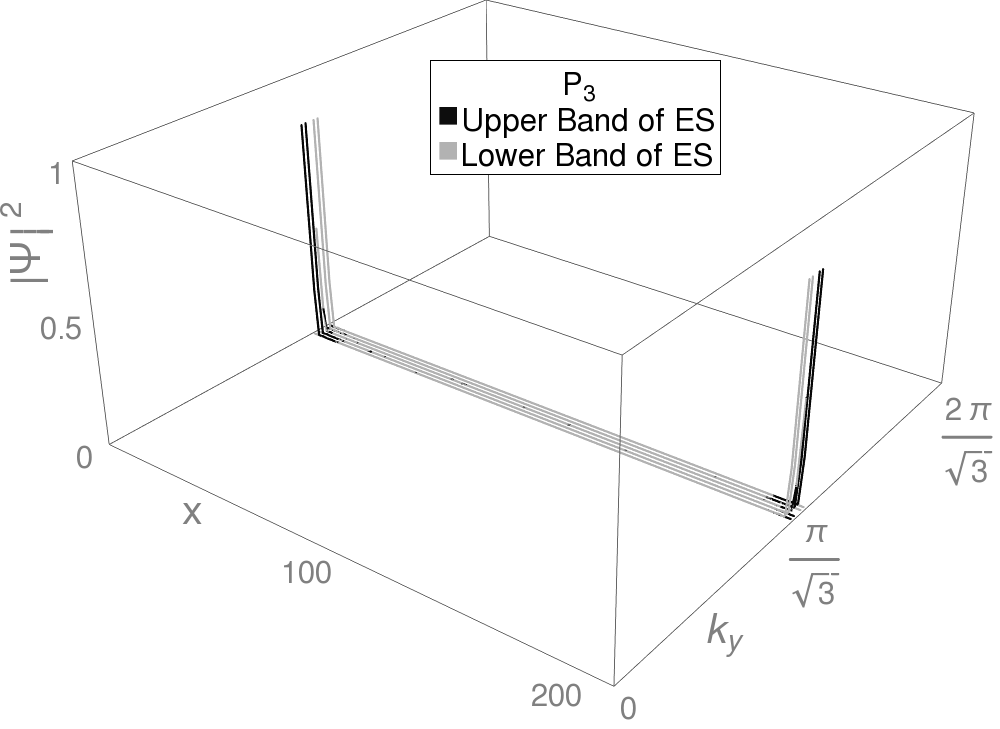}}\\
	\subfloat[$P_4=\left(A_0a,\Omega/t_h\right)=\left(10,.5\right),C=0$]{\includegraphics[width = .9\columnwidth]{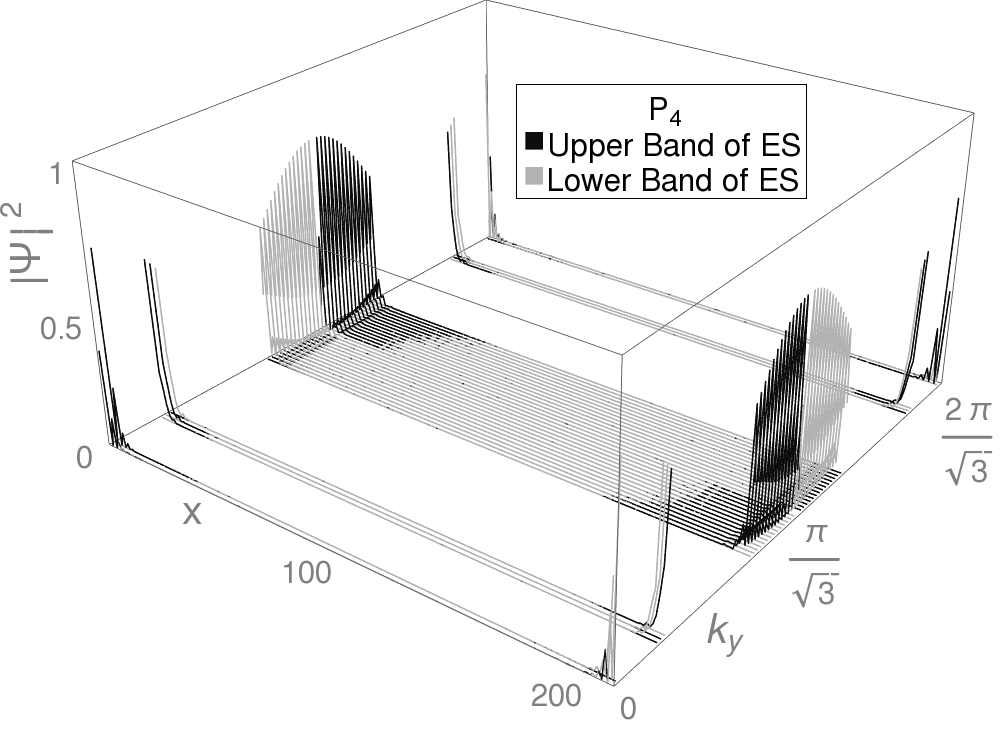}}
\end{tabular}
\caption{(Color online) Edge modes of the Floquet state for the phases $P_{3,4}$. Upper (lower) band corresponds to entanglement eigenvalues
$\epsilon \geq 1/2$ ($\epsilon \leq 1/2$). Only states corresponding to $\epsilon$ within a small window of $1/2$ are shown, to highlight those that
disperse through $1/2$. The $P_3$ phase contains a left mover on $x=0$ edge and a right mover on the $x = 200$ edge. 
The $P_4$ phase contains two left movers and two right movers on both edges. The varying ``width" of the edge states in the $k_y$ label is 
indicative of the slope of the crossing bands.}
\label{fig:EdgeF34}
\end{figure}

\begin{figure}
\begin{tabular}{c}
	\subfloat[$P_1=\left(A_0a,\Omega/t_h\right)=\left(.5,10\right),C=1$]{\includegraphics[width = .9\columnwidth]{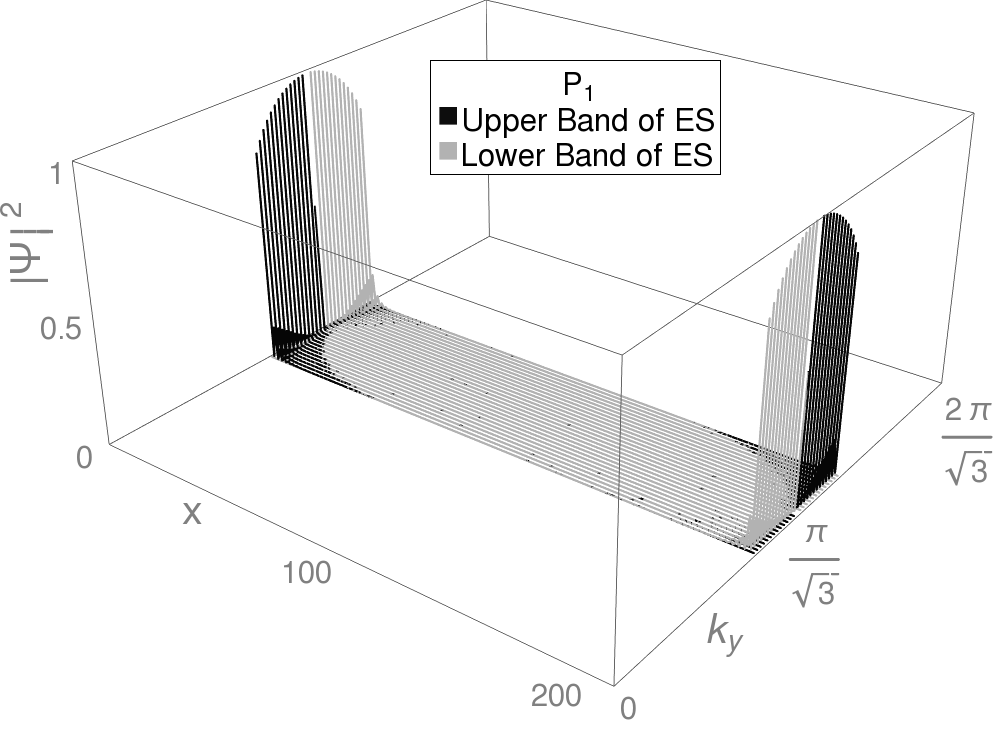}}\\
	\subfloat[$P_2=\left(A_0a,\Omega/t_h\right)=\left(.5,5\right),C=3$]{\includegraphics[width = .9\columnwidth]{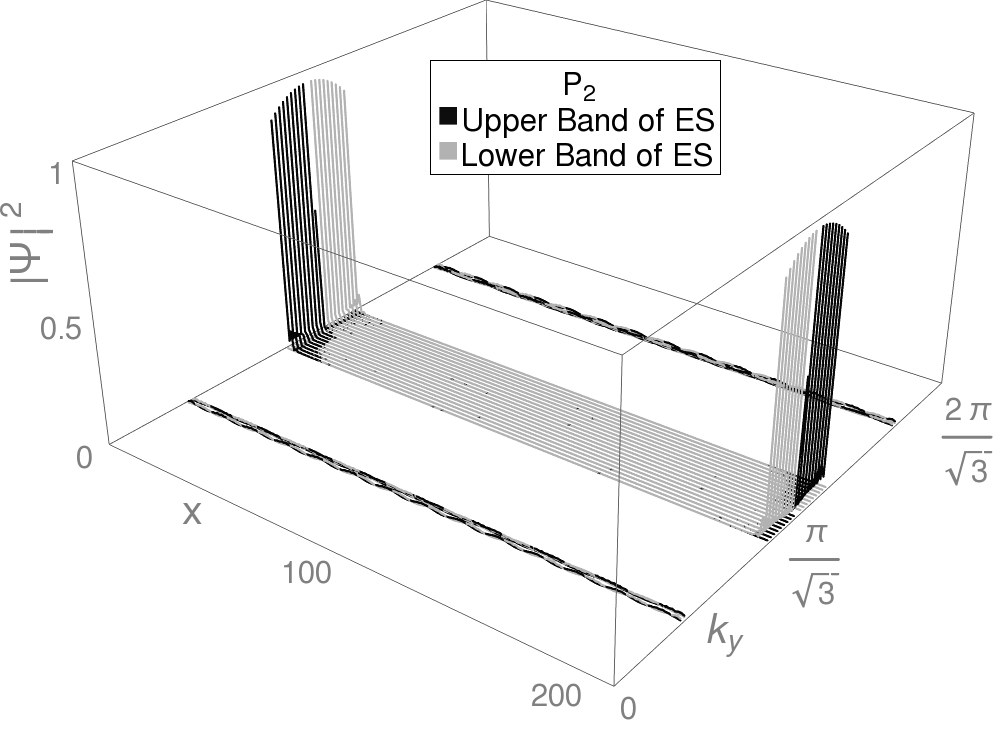}}\\
\end{tabular}
\caption{(Color online) Edge modes of the quench state for the phases $P_{1,2}$. Note the absence of edge states in phase $P_2$
for $k_y=\pi(1\pm 0.74)/\sqrt{3}$. In both phases we see a left mover on $x=0$ edge and a right mover on the $x=200$ edge. }
\label{fig:EdgeQ12}
\end{figure}

\begin{figure}
\begin{tabular}{c}
%	\subfloat[
%$P_3=\left(A_0a,\Omega/t_h\right)=\left(1.5,5\right),C=1$
%]
{\includegraphics[width = .9\columnwidth]{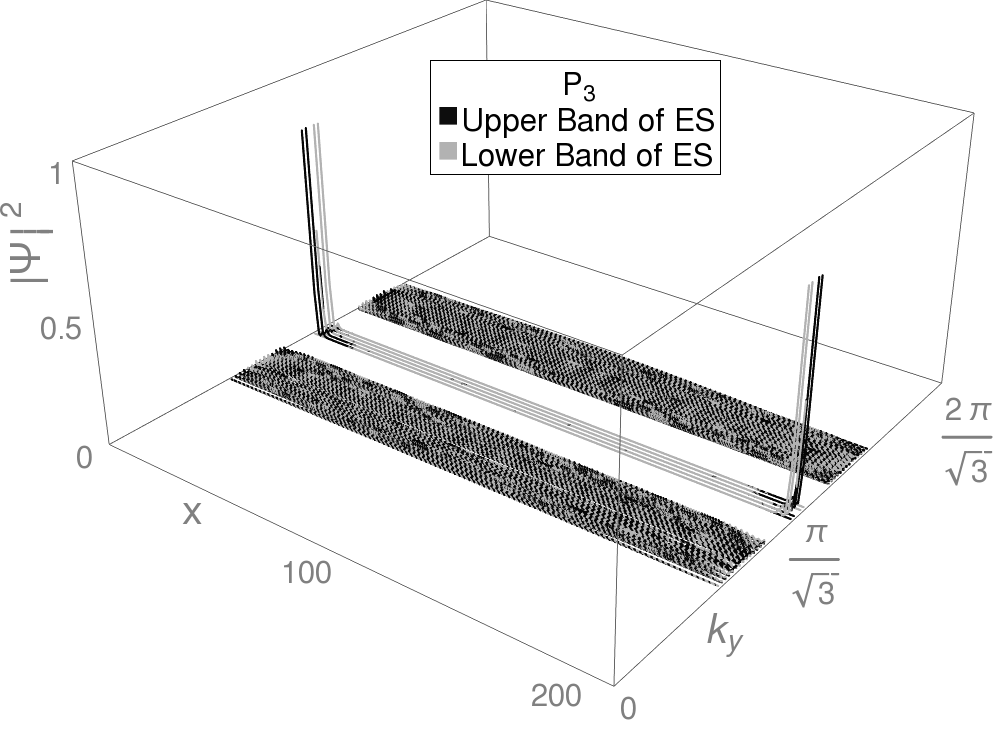}}
\end{tabular}
\caption{(Color online) Edge modes of the quench state for the phase $P_3$. A left mover resides on the $x=0$ edge and a right mover is localized on the $x= 200$ 
edge. The delocalized states correspond to bulk excitations surrounding the edge mode.}
\label{fig:EdgeQ34}
\end{figure}

The amplitudes of the edge states are shown in Figures~\ref{fig:EdgeF12},~\ref{fig:EdgeF34} for the Floquet ground state,
and in Figures~\ref{fig:EdgeQ12},~\ref{fig:EdgeQ34} for the true state. 
To be clear, by ``state" we mean the eigenvector of the correlation matrix $\hat{C}$. As discussed in the main text, we expect these edge states,
at least for the Floquet ground state, to behave as the usual edge states of a topological Hamiltonian.
In fact, the states are highly localized to the edge with a decay length
on the order of one lattice site.  Moreover, we can directly verify the chirality of the edge states by analyzing how the wavefunctions vary with
momentum $k_y$.

In Figures~\ref{fig:EdgeF12},~\ref{fig:EdgeF34},~\ref{fig:EdgeQ12},~\ref{fig:EdgeQ34}
the two states with $\epsilon$ closest to $1/2$ are shown.
The higher $\epsilon\geq 1/2$ state - analogous to the higher energy state - is colored dark while the lower state ($\epsilon\leq 1/2$) is colored light.
As can be seen at $k_y$ near the crossing point, the edge states are well defined and localized on opposite edges. Moreover, at the crossing,
the high $\epsilon$ state switches with its counterpart. This is expected in a chiral edge because the two bands cross as $k_y$ is varied.
Therefore, by analyzing the pattern of switching, the chirality of the bands stated in the main text can be verified.

As an example of this analysis we consider the $x= 0$ edge in Figure~\ref{fig:EdgeF12} for the FGS of phase $P_2$. 
We see a pattern of black to gray for all three crossings, which indicates that all the downward arching bands in Figure~\ref{fig4} 
correspond to chiral edges that reside on the $x=0$ edge of the strip. The edge states corresponding to the downward trending bands at each crossing will 
be termed ``left movers" and the upward arching bands will be termed ``right-movers". Thus, by the same analysis,
the $x=200$ edge is found to contain all right movers at each crossing.
 
The same can be done for the FGS of the $P_4$ phase to verify the assignment of $C = 0$. Both sides contain two crossings of both types, black
to gray and gray to black. Thus each side contains an equal number of left movers and right movers.  Note that these edge states 
vanish altogether at other times during the 
laser period, as pointed out in Figure~\ref{fig:P4Ent}. In contrast, for all other phases with $C\neq 0$, the edge-states  
slightly shift their $k_y$ positioning in a periodic manner. 

Figures~\ref{fig:EdgeQ12},~\ref{fig:EdgeQ34} contain states corresponding to $\epsilon$ near $1/2$ for the unitarily evolved system. These figures show that
unlike the FGS, not all the states at this eigenvalue are edge states. While phase $P_1$ is similar to the FGS, for phases $P_{2,3}$ only one pair of 
edge modes exist, with a large fraction of states around $\epsilon\sim 1/2$ being bulk states.

\begin{figure}
\begin{center}
\includegraphics[width = 1.0\columnwidth]{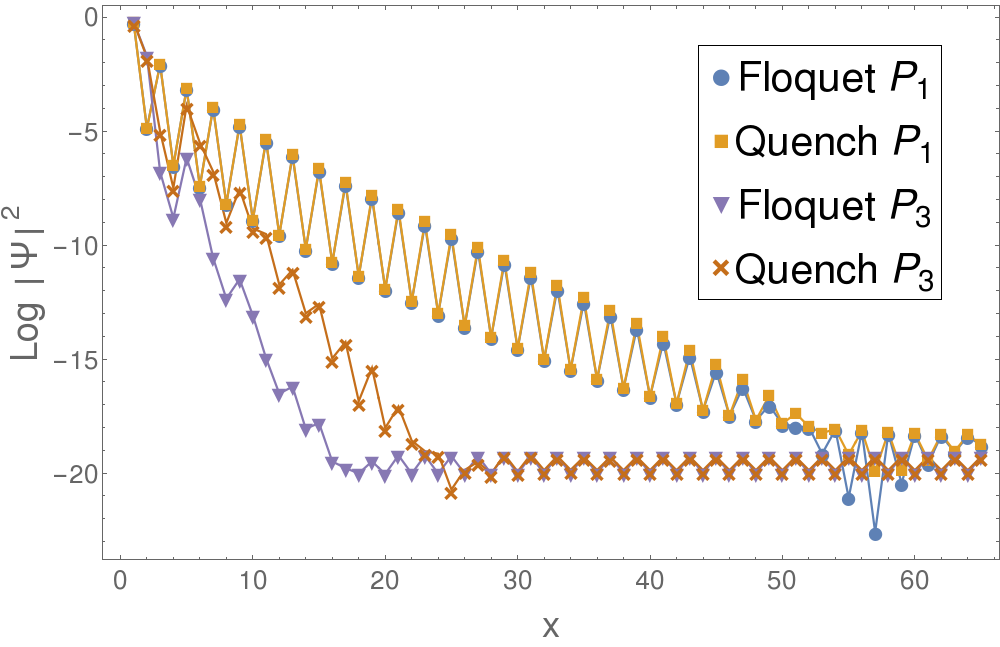}
\caption{(Color online) Decay of the edge states in the ES shown as the logarithm of the mod-squared of the wavefunction against the distance from the edge.
The states shown correspond to the $\epsilon = 1/2$ state at $k_y= \pi/\sqrt{3}$, the ``zero energy'' state of the conventional edge mode.
Circles - $P_1$ Floquet state, squares - $P_1$ quench state, triangles - $P_3$ Floquet state, crosses - $P_3$ quench state.
Even $x$ are $A$ sites and odd $x$ are $B$ sites so the apparent rapid oscillation is simply due to a different weight on the $A$ and $B$ sublattices.
In all cases the decay length is on the order of one lattice site.  \label{fig:edgedecay}}
\end{center}
\end{figure}

The decay of the edge states is shown in Figure~\ref{fig:edgedecay}. The decay lengths of the edge states are determined by the inverse of the entanglement
gap, which is controlled by the bulk excitation density. Thus $P_1$ which has few bulk excitations in the quenched state has a similar decay length 
to the FGS. In contrast, since the quenched state of $P_3$ has more 
bulk excitations, it shows a longer decay length than the edge state of the FGS. The surviving edge state of $P_2$ in the quenched state 
has a similar decay length as the FGS because, as seen in Fig.~\ref{fig4}, in the vicinity of this edge state the entanglement gap of the quenched state is almost the
same as that of the FGS.

%############### 3D PLOTS ###########################

%############### PROJECTION PLOTS ###########################

%############### DECAY PLOTS, MERGED (HIGHER QUALITY) ###########################

%############### 3D EDGE MODE PLOTS ###########################

%################DELTA RHO vs. DISPERSION###################
\begin{figure}
\begin{center}
\includegraphics[width = 1.0\columnwidth]{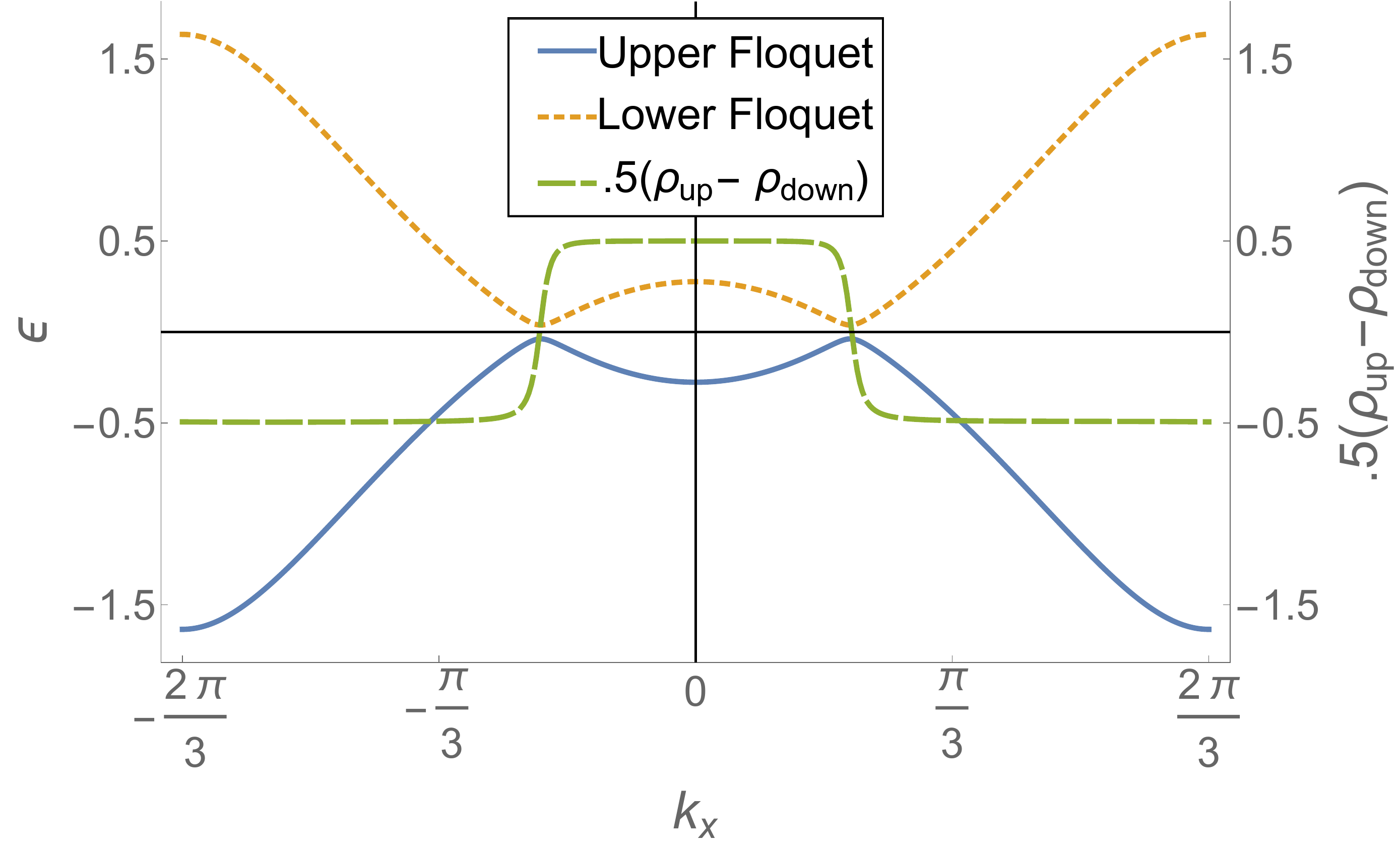}
\end{center}
\caption{(Color online) Population along $k_x$, for $k_y=\frac{\pi}{\sqrt{3}}(1 - .74)$ where for
$P_2=\left(A_0a,\frac{\Omega}{t_h}\right)=\left(0.5,5\right), C=3$ an edge mode appears in the ES of the Floquet eigenstate,
but vanishes in the ES of the true
state.  Also plotted are the dispersions for the upper and lower Floquet bands. Note that $2\delta\rho_k=\rho_{k,\rm down}-\rho_{k,\rm up}$
goes through zero precisely at the
closest approach of the two bands, as expected of a weak resonant laser.}
\label{fig13}
\end{figure}

\section{Derivation of Eq.~(\ref{eq:Cinf})}\label{app4}
Generally, the correlation matrix is given by
\begin{align}
\hat{C} &
		= \int_{BZ} \!\!\!d^2 k e^{i k\cdot r} \Bigg[|\psi_{ka}|^2 |a_k(t)\rangle\langle a_k(t)|
		+ |\psi_{kb}|^2 |b_k(t)\rangle\langle b_k(t)| \nonumber\\
		& + e^{-i\left(\epsilon_{ka}-\epsilon_{kb}\right)t}\psi_{ka}\psi_{kb}^*|a_k(t)\rangle\langle b_k(t)|\nonumber\\
&+ e^{i\left(\epsilon_{ka}-\epsilon_{kb}\right)t}\psi_{ka}^*\psi_{kb}|b_k(t)\rangle\langle a_k(t)|\Bigg],
\end{align}
where $\psi_{ka} = \langle a_k(0)  | \psi_{k,\rm in}\rangle$ and likewise for $\psi_{kb}$, and $\psi_{k,\rm in}$ is the initial state.
We make the assumption that the third and fourth off-diagonal terms proportional to the exponential 
vanish at long times because of the rapidly varying phase.
We emphasize that this is an acceptable assumption because we are considering $r$ to be restricted to a sub region of the lattice.
For general $r$ the stationary phase approximation implies that the exponentials will contribute in a region where $ r \approx v_k t$
where $v_k$ is the group velocity of the Floquet dispersion.  Therefore, as long as we consider times that are much larger than $r/v$, where $v$ is
some characteristic velocity, we may neglect the oscillating terms.
We have also checked the validity of our argument from the numerical results.  We should mention that the order of limits here does not commute.
The limit $t\rightarrow\infty$ must be taken before width $\rightarrow\infty$. Physically this corresponds to only analyzing the interior of the light-cone.

If we neglect these off-diagonal terms, we obtain
\begin{align}
\hat{C} &
		= \int_{BZ} \!\!\!d^2 k e^{i k\cdot r} \Bigg[|\psi_{ka}|^2 |a_k(t)\rangle\langle a_k(t)|\nonumber\\
&		+ |\psi_{kb}|^2 |b_k(t)\rangle\langle b_k(t)| \Bigg].
\end{align}

Now employing $|a_k(t)\rangle\langle a_k(t)| + |b_k(t)\rangle\langle b_k(t)| = 1$
from the completeness of the basis, and the fact that $|\psi_{ka}|^2 + |\psi_{kb}|^2 = 1$, we obtain the stated equation where
we identify $\rho_{k,\rm down}= |\psi_{ka}|^2, \rho_{k,\rm up}= |\psi_{kb}|^2$.

\section{Edge states in the entanglement spectrum of the Dirac model}\label{app5}
Before we discuss the ES, let us review how edge states appear in the physical boundaries of the Dirac Hamiltonian.
The Dirac Hamiltonian in a semi-infinite geometry with a boundary at $x=0$ is,
\begin{eqnarray}
H= v_F\biggl[-i\sigma_x\partial_x -i \sigma_y\partial_y\biggr] + m(x)\sigma_z.
\end{eqnarray}
The boundary is controlled by the behavior of $m(x)$.
We may write the eigenstate
as $\Psi_k(x,y) = e^{i k_y y}e^{-\frac{1}{v_F}\int^x m(x') dx'}\chi$, which should obey
\begin{eqnarray}
&&\biggl(v_F\biggl[-i\sigma_x\partial_x -i \sigma_y\partial_y\biggr] + m(x)\sigma_z\biggr)\Psi_k(x,y) \nonumber\\
&&= E_k\Psi_k(x,y),
\end{eqnarray}
and implies the eigenvalue equation,
\begin{eqnarray}
m(x)\biggl[i \sigma_x  + \sigma_z \biggr]\chi + v_F k_y \sigma_y \chi   = E_k \chi.
\end{eqnarray}
The solution is $E_k= v_F k_y$, $\sigma_y \chi=\chi$.
Thus the eigenstate is,
\begin{eqnarray}
\Psi_k(x,y) = e^{i k_y y}e^{-\frac{1}{v_F}\int^x m(x') dx'}\frac{1}{\sqrt{2}}\begin{pmatrix}1\\ i\end{pmatrix}.
\end{eqnarray}
The above is an edge state with a decay length controlled by $v_F/m$.

Now let us consider how edge states emerge in the entanglement spectrum. For this, we will consider a spatially uniform system ($m(x)=m$)
with eigenstates,
\begin{eqnarray}
&&|\phi_{k\pm}\rangle= \frac{1}{\sqrt{v_F^2 k^2 + \left(E_{k\pm}-m\right)^2}}\begin{pmatrix}v_F(k_x-ik_y) \\ E_{k\pm} -m \end{pmatrix},\\
&&E_{k\pm}= \pm \sqrt{v_F^2 k^2 + m^2}.
\end{eqnarray}
It is convenient to define
\begin{eqnarray}
&&C_{\pm}(k)= |\phi_{k\pm}\rangle\langle\phi_{k\pm}|,
\end{eqnarray}
where the correlation function in real space is
\begin{eqnarray}
&&C_{\pm}(\vec{r}) = \int \frac{d^2k}{(2\pi)^2} e^{i k_x r_x + i k_y r_y} C_{\pm}(k).
\end{eqnarray}
For any generic state, be it one generated at long times after a quench when dephasing has set in, or a mixed state at finite temperature,
we may write,
\begin{eqnarray}
&&C_{\rm quench}(\vec{r}) \!\!= \int \frac{d^2k}{(2\pi)^2} \sum_{s=\pm}e^{i k_x r_x + i k_y r_y} \rho_{s}(k)C_{s}(k).
\end{eqnarray}
We will assume the system to be at half-filling.
Note that since, $C_+(k) + C_-(k) = 1$, a filled band will give,
\begin{eqnarray}
&&C_+(r) + C_-(r) =  \int \frac{d^2k}{(2\pi)^2} e^{i k_x r_x + i k_y r_y}\begin{pmatrix}1 & 0 \\0 & 1 \end{pmatrix},\nonumber\\
&&=\delta^2(r).
\end{eqnarray}
The above corresponds to an unentangled product state with eigenvalues of the correlation matrix being $1$.

It is useful to note that
\begin{eqnarray}
&&C_+(k) - C_-(k) =\nonumber\\
&&\frac{1}{\sqrt{v_F^2k^2 + m^2}}\begin{pmatrix}m &v_F(k_x-ik_y)\\ v_F(k_x+ik_y)
& -m \end{pmatrix},
\end{eqnarray}
is a matrix with eigenvalues $\pm 1$.
We write $\rho_{\pm}(k) = 1/2 \pm \delta \rho_k$, where $\delta \rho_k=\pm 1/2$ are pure states. Since
we are interested in the ES of a semi-infinite geometry ($x>0$), we may study the $k_y$ Fourier component
of $C$ in the translationally invariant $y$ direction,
\begin{eqnarray}
&&C_{\rm quench}(x,k_y) = \frac{\delta (x)}{2}
+ \int \frac{dk_x}{2\pi}\frac{\delta \rho_k}{\sqrt{v_F^2k_x^2 +v_F^2 k_y^2 +m^2}}\nonumber\\
&&\times e^{ik_x x}
\begin{pmatrix}m &v_F(k_x-ik_y)\\ v_F(k_x+ik_y)&-m\end{pmatrix}.\nonumber\\
\end{eqnarray}
Note that the above correlation matrix is $\left(1/2\right)\delta(r)$ added to the Fourier transform of
$\delta\rho_{k}$ times a matrix with eigenvalues $\pm 1$. Therefore, in the infinite system limit it has eigenvalues
$1/2 \pm \delta\rho_k$ corresponding to plane-waves eigenvectors.
In the presence of a boundary we expect some linear combinations of these plane waves to form extended scattering states.
The boundary is now determined by the entanglement cut at $x=0$, rather than $m(x)$.

To determine the edge states, it is convenient to define the integral
\begin{eqnarray}
\int \frac{dk_x}{2\pi}\frac{\delta \rho_{k_y,k_x}}{\sqrt{v_F^2k_x^2 +v_F^2 k_y^2 +m^2}}e^{ik_x x}\!\!= \!I\left(x,k_y,m\right).
\end{eqnarray}
Then generically,
\begin{eqnarray}
&&C_{\rm quench}(x,k_y) = \frac{\delta (x)}{2} \nonumber\\
&&+ \biggl(m \sigma_z + v_F k_y\sigma_y -i v_F\sigma_x \partial_x\biggr)I\left(x,k_y,m\right).
\end{eqnarray}

If $\delta \rho_k\neq 0$ and
smoothly varying around $k_x=0$, we may approximate it by it's value at $k_x = 0$. This may then be pulled out of the integral.
Then using the fact that
\begin{eqnarray}
&&\int \frac{dk_x}{2\pi}\frac{e^{ik_x x}}{\sqrt{v_F^2k_x^2 +v_F^2 k_y^2 +m^2}}\nonumber\\
&&= \frac{1}{\pi v_F}
K_0\left(\frac{|x|\sqrt{m^2 + v_F^2 k_y^2} }{v_F}\right),\nonumber
\end{eqnarray}
and since $K_0(x)$ decays exponentially away from $x=0$, this term can be replaced by a delta-function, $K_0(|x|) \simeq \pi \delta(x)$.
Thus,
\begin{eqnarray}
&&C_{\rm quench}(x,k_y)\simeq \delta (x) \biggl[\frac{1}{2} \nonumber\\
&&+ \frac{\delta \rho_{k_y,k_x=0}}{\sqrt{m^2 + v_F^2 k_y^2}}
\biggl(m \sigma_z + v_F k_y\sigma_y + i v_F\sigma_x \partial_x\biggr)
\biggr],
\end{eqnarray}
up to terms which are higher order in derivatives.
Note that if $\delta \rho_{k_y,k_x = 0} = 0$ the second term vanishes and $C$ is controlled by the higher order terms which will 
not generically have edge states.

The entanglement eigenvalues are determined by solving
\begin{eqnarray}
\int dx' C(x,x';k_y)\Psi_{k_y}(x') = \epsilon_{k_y}\Psi(x).
\end{eqnarray}
This implies that
\begin{eqnarray}
 &&\biggl[\frac{1}{2} + \frac{\delta \rho_{k_y,k_x=0}}{\sqrt{m^2 + v_F^2 k_y^2}}
\biggl(m \sigma_z + v_F k_y\sigma_y + i v_F\sigma_x \partial_x\biggr)
\biggr]\Psi_{k_y}(x) \nonumber\\
&&= \epsilon_{k_y} \Psi_{k_y}(x).
\end{eqnarray}

This is solved by the same ansatz as for the usual physical edge mode of the Dirac Hamiltonian, 
$\Psi_{k_y}(x) = \theta (x)e^{i k_y y} e^{-\frac{m x}{v_F}}\chi$ where, $\sigma_y\chi = -\chi$ and the step 
function to explicitly impose the boundary conditions.
The entanglement eigenvalues are,
\begin{eqnarray}
\epsilon_{k_y} = \frac{1}{2} - \delta \rho_{k_y,k_x=0}\frac{v_F k_y}{\sqrt{m^2 + v_F^2 k_y^2}}.
\end{eqnarray}
In particular this shows that the edge state with $\epsilon = 1/2$ exists at $k_y = 0$, and it disperses linearly with $k_y$ about this point.

Now we return to the assumption we made that $\delta\rho_k \neq 0$.
If  $\delta\rho_{k} =0$ were to hold at some point in $k_x$, then there would be an extended state with $\epsilon_{k_y} = 1/2$.
As this extended state, and edge mode, both have equal $k_y$ momentum, and the same eigenvalue under $C$, we expect them to mix.
Therefore, in this case expanding around $k_x = 0$ is not valid, the argument above does not hold, and we do not expect edge states.

While this analytic argument was shown for the Dirac model, and the Floquet Chern insulator is far more complicated, especially
for large Chern numbers, yet we find that the same principles apply. For example Figure~\ref{fig13} shows how $\rho_{k_x,k_y^*}$
varies along $k_x$ for the $P_2$ phase. We have chosen $k_y^*=\pi (1-0.74)/\sqrt{3}$ which coincides with the edge mode in the ES
of the Floquet eigenstate that is located
to the left of the central edge mode  (see Figure \ref{fig4} (b) in main text). This particular edge mode is absent in the true wavefunction.
The reason now is clear.
Figure \ref{fig13} shows that $\delta \rho_{k_x,k_y^*}$ vanishes at two points in $k_x$, implying extended states with the same
energy as the edge state. These extended and edge states can now mix, resulting in the absence of the edge state at
$k_y^*$. It is interesting
to note that the resonant laser results in a rather precise location of the vanishing of
$\delta\rho_k$ relative to the avoided band crossings of the upper and lower Floquet bands.

\end{appendix}

%\bibliography{quench}


\end{document}